\documentclass[aps,prd,twocolumn,superscriptaddress,amsmath,amssymb,nofootinbib]{revtex4-1}
\pdfoutput=1
\usepackage{graphicx}
\usepackage{dcolumn}
\usepackage{bm}
\usepackage{hyperref}

\begin{document}
\title{Optimizing EDELWEISS detectors for low-mass WIMP searches}
\author{Q.~Arnaud}
\altaffiliation[Now at ]{Department of Physics, Queen’s University, Kingston, K7L 3N6, Canada; }
\email{q.arnaud@queensu.ca}
\affiliation{Univ Lyon, Universit\'{e} Lyon 1, CNRS/IN2P3, IPN-Lyon, F-69622, Villeurbanne, France}
\author{E.~Armengaud}
\affiliation{IRFU, CEA, Universit\'{e} Paris-Saclay, F-91191 Gif-sur-Yvette, France}
\author{C.~Augier}
\affiliation{Univ Lyon, Universit\'{e} Lyon 1, CNRS/IN2P3, IPN-Lyon, F-69622, Villeurbanne, France}
\author{A.~Beno\^{i}t}
\affiliation{Institut N\'{e}el, CNRS/UJF, 25 rue des Martyrs, BP 166, 38042 Grenoble, France}
\author{L.~Berg\'{e}}
\affiliation{CSNSM, Univ. Paris-Sud, CNRS/IN2P3, Universit\'{e} Paris-Saclay, 91405 Orsay, France}
\author{J.~Billard}
\affiliation{Univ Lyon, Universit\'{e} Lyon 1, CNRS/IN2P3, IPN-Lyon, F-69622, Villeurbanne, France}
\author{A.~Broniatowski}
\affiliation{CSNSM, Univ. Paris-Sud, CNRS/IN2P3, Universit\'{e} Paris-Saclay, 91405 Orsay, France}
\author{P.~Camus}
\affiliation{Institut N\'{e}el, CNRS/UJF, 25 rue des Martyrs, BP 166, 38042 Grenoble, France}
\author{A.~Cazes}
\affiliation{Univ Lyon, Universit\'{e} Lyon 1, CNRS/IN2P3, IPN-Lyon, F-69622, Villeurbanne, France}
\author{M.~Chapellier}
\affiliation{CSNSM, Univ. Paris-Sud, CNRS/IN2P3, Universit\'{e} Paris-Saclay, 91405 Orsay, France}
\author{F.~Charlieux}
\affiliation{Univ Lyon, Universit\'{e} Lyon 1, CNRS/IN2P3, IPN-Lyon, F-69622, Villeurbanne, France}
\author{M. De~J\'{e}sus}
\affiliation{Univ Lyon, Universit\'{e} Lyon 1, CNRS/IN2P3, IPN-Lyon, F-69622, Villeurbanne, France}
\author{L.~Dumoulin}
\affiliation{CSNSM, Univ. Paris-Sud, CNRS/IN2P3, Universit\'{e} Paris-Saclay, 91405 Orsay, France}
\author{K.~Eitel}
\affiliation{Karlsruher Institut f\"{u}r Technologie, Institut f\"{u}r Kernphysik, Postfach 3640, 76021 Karlsruhe, Germany}
\author{N.~Foerster}
\affiliation{Karlsruher Institut f\"{u}r Technologie, Institut f\"{u}r Experimentelle Kernphysik, Gaedestr. 1, 76128 Karlsruhe, Germany}
\author{J.~Gascon}
\affiliation{Univ Lyon, Universit\'{e} Lyon 1, CNRS/IN2P3, IPN-Lyon, F-69622, Villeurbanne, France}
\author{A.~Giuliani}
\affiliation{CSNSM, Univ. Paris-Sud, CNRS/IN2P3, Universit\'{e} Paris-Saclay, 91405 Orsay, France}
\author{M.~Gros}
\affiliation{IRFU, CEA, Universit\'{e} Paris-Saclay, F-91191 Gif-sur-Yvette, France}
\author{L.~Hehn}
\altaffiliation[Now at ]{Nuclear Science Division, Lawrence Berkeley National Laboratory, Berkeley, CA, US; }
\affiliation{Karlsruher Institut f\"{u}r Technologie, Institut f\"{u}r Kernphysik, Postfach 3640, 76021 Karlsruhe, Germany}
\author{Y.~Jin}
\affiliation{Laboratoire de Photonique et de Nanostructures, CNRS, Route de Nozay, 91460 Marcoussis, France}
\author{A.~Juillard}
\affiliation{Univ Lyon, Universit\'{e} Lyon 1, CNRS/IN2P3, IPN-Lyon, F-69622, Villeurbanne, France}
\author{M.~Kleifges}
\affiliation{Karlsruher Institut f\"{u}r Technologie, Institut f\"{u}r Prozessdatenverarbeitung und Elektronik, Postfach 3640, 76021 Karlsruhe, Germany}
\author{V.~Kozlov}
\affiliation{Karlsruher Institut f\"{u}r Technologie, Institut f\"{u}r Experimentelle Kernphysik, Gaedestr. 1, 76128 Karlsruhe, Germany}
\author{H.~Kraus}
\affiliation{University of Oxford, Department of Physics, Keble Road, Oxford OX1 3RH, UK}
\author{V. A.~Kudryavtsev}
\affiliation{University of Sheffield, Department of Physics and Astronomy, Sheffield, S3 7RH, UK}
\author{H.~Le-Sueur}
\affiliation{CSNSM, Univ. Paris-Sud, CNRS/IN2P3, Universit\'{e} Paris-Saclay, 91405 Orsay, France}
\author{R.~Maisonobe}
\affiliation{Univ Lyon, Universit\'{e} Lyon 1, CNRS/IN2P3, IPN-Lyon, F-69622, Villeurbanne, France}
\author{S.~Marnieros}
\affiliation{CSNSM, Univ. Paris-Sud, CNRS/IN2P3, Universit\'{e} Paris-Saclay, 91405 Orsay, France}
\author{X.-F.~Navick}
\affiliation{IRFU, CEA, Universit\'{e} Paris-Saclay, F-91191 Gif-sur-Yvette, France}
\author{C.~Nones}
\affiliation{IRFU, CEA, Universit\'{e} Paris-Saclay, F-91191 Gif-sur-Yvette, France}
\author{E.~Olivieri}
\affiliation{CSNSM, Univ. Paris-Sud, CNRS/IN2P3, Universit\'{e} Paris-Saclay, 91405 Orsay, France}
\author{P.~Pari}
\affiliation{IRAMIS, CEA, Universit\'{e} Paris-Saclay, F-91191 Gif-sur-Yvette, France}
\author{B.~Paul}
\affiliation{IRFU, CEA, Universit\'{e} Paris-Saclay, F-91191 Gif-sur-Yvette, France}
\author{D.~Poda}
\affiliation{CSNSM, Univ. Paris-Sud, CNRS/IN2P3, Universit\'{e} Paris-Saclay, 91405 Orsay, France}
\author{E.~Queguiner}
\affiliation{Univ Lyon, Universit\'{e} Lyon 1, CNRS/IN2P3, IPN-Lyon, F-69622, Villeurbanne, France}
\author{S.~Rozov}
\affiliation{JINR, Laboratory of Nuclear Problems, Joliot-Curie 6, 141980 Dubna, Moscow Region, Russian Federation}
\author{V.~Sanglard}
\affiliation{Univ Lyon, Universit\'{e} Lyon 1, CNRS/IN2P3, IPN-Lyon, F-69622, Villeurbanne, France}
\author{S.~Scorza}
\altaffiliation[Now at ]{SNOLAB, Lively, ON, Canada; }
\affiliation{Karlsruher Institut f\"{u}r Technologie, Institut f\"{u}r Experimentelle Kernphysik, Gaedestr. 1, 76128 Karlsruhe, Germany}
\author{B.~Siebenborn}
\affiliation{Karlsruher Institut f\"{u}r Technologie, Institut f\"{u}r Kernphysik, Postfach 3640, 76021 Karlsruhe, Germany}
\author{L.~Vagneron}
\affiliation{Univ Lyon, Universit\'{e} Lyon 1, CNRS/IN2P3, IPN-Lyon, F-69622, Villeurbanne, France}
\author{M.~Weber}
\affiliation{Karlsruher Institut f\"{u}r Technologie, Institut f\"{u}r Prozessdatenverarbeitung und Elektronik, Postfach 3640, 76021 Karlsruhe, Germany}
\author{E.~Yakushev}
\affiliation{JINR, Laboratory of Nuclear Problems, Joliot-Curie 6, 141980 Dubna, Moscow Region, Russian Federation}
\collaboration{The EDELWEISS Collaboration}


\begin{abstract}
The physics potential of EDELWEISS detectors for the search of low-mass Weakly Interacting Massive Particles (WIMPs) is studied.
Using a data-driven background model, projected exclusion limits are computed using frequentist and multivariate analysis approaches, namely profile likelihood and boosted decision tree.
Both current and achievable experimental performance are considered. 
The optimal strategy for detector optimization depends critically on whether the emphasis is put on WIMP masses below or above $\sim 5$~GeV/c$^2$. 
The projected sensitivity for the next phase of the EDELWEISS-III experiment at the Modane Underground Laboratory (LSM) for low-mass WIMP search is presented. By 2018 an upper limit on the spin-independent WIMP-nucleon cross-section of $\sigma_{SI} = 7 \times 10^{-42}$~cm$^2$ is expected for a WIMP mass in the range $2-5$~GeV/c$^2$.
The requirements for a future hundred-kilogram scale experiment designed to reach the bounds imposed by the coherent scattering of solar neutrinos are also described. By improving the ionization resolution down to 50 eV$_{ee}$, we show that such an experiment installed in an even lower background environment (e.g. at SNOLAB) should allow to observe about 80 $^8$B neutrino events after discrimination.

\end{abstract}
\maketitle
\section{Introduction}
\begin{figure*}[htb]
\includegraphics[width=1.1\textwidth]{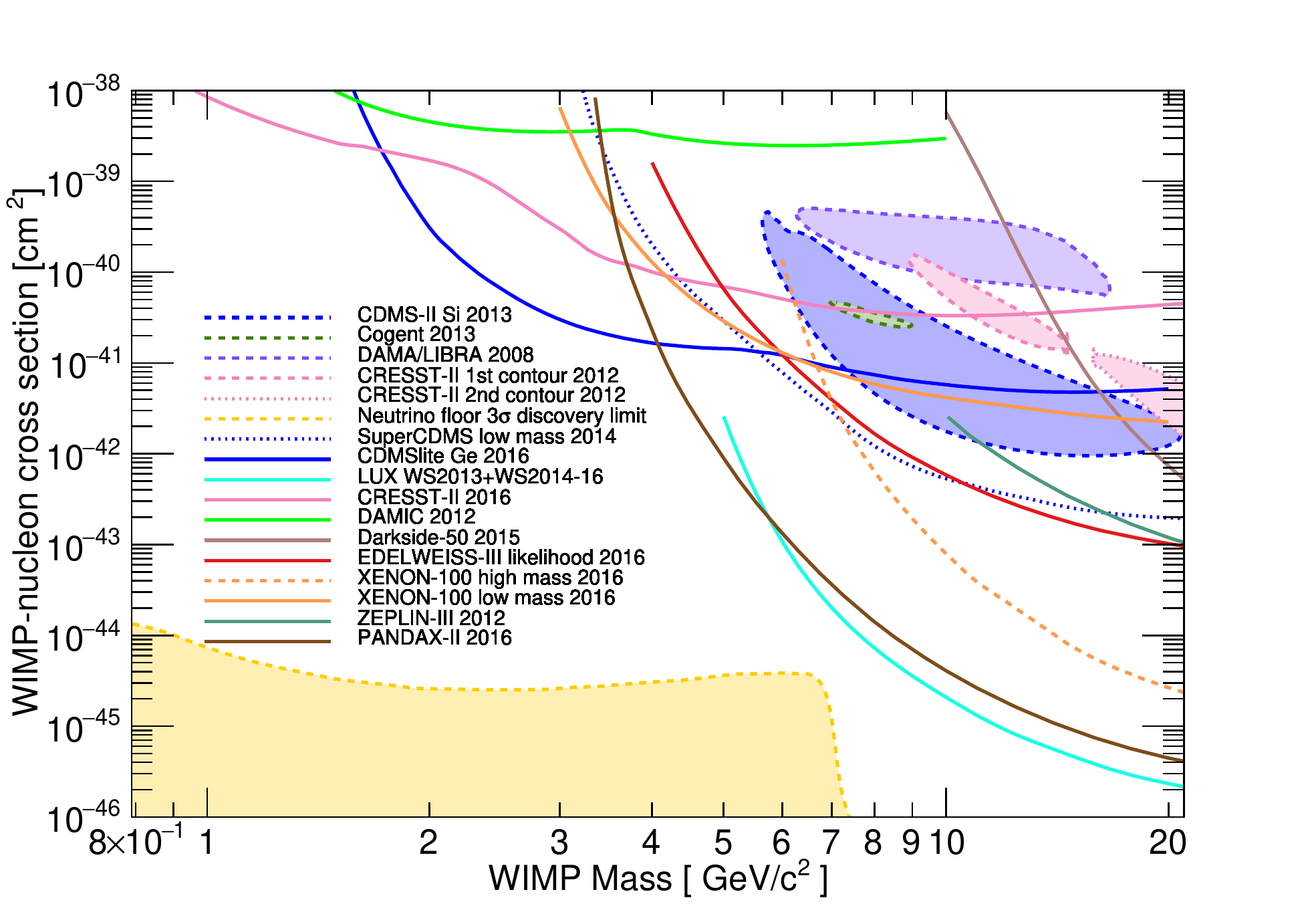}
\caption{\label{fig:SotA2016}Constraints in the spin-independent (SI) WIMP-nucleon cross section as a function of WIMP mass. Closed contours corresponding to signal hints reported by CDMS-II Si~\cite{cdmsSI-contour} (dashed blue, 90\% C.L.), CoGeNT~\cite{CoGeNT-contour} (dashed green, 90\% C.L.), CRESST-II~\cite{CRESST-contour} (dashed  pink, 95\% C.L.), and DAMA/LIBRA~\cite{DAMA-contour} (dashed purple,  90\% C.L.) experiments.  Results interpreted as 90\% C.L. exclusion upper limits are represented by lines:  experimental limits shown are from DAMIC~\cite{DAMIC} (green), CRESST~\cite{CRESST} (pink), SuperCDMS low mass~\cite{SuperCDMS2014} (dotted dark blue), CDMSlite with Ge~\cite{CDMSlite2015} (dark blue), PandaX-II~\cite{PandaX-II-2016} (brown), LUX combined~\cite{Lux2016-2} (turquoise blue), XENON-100 high- and low-mass~\cite{xenon100-2016} (dashed orange and orange), EDELWEISS low mass~\cite{EDWlikelihood} (red), ZEPLIN-III~\cite{Zeplin} (blue-green) and DarkSide-50~\cite{DarkSide50} (pale brown). The region delimited by the yellow dashed line corresponds to the neutrino floor~\cite{Billard}.}
\end{figure*}

In the past decades, astronomical surveys and cosmological precision measurements have led to the worldwide consensus that the matter content of the Universe is dominated by non-baryonic dark matter~\cite{Planck2015}. Though its nature remains unknown, a class of dark matter candidates from physics beyond the Standard Model is so far favoured and known as Weakly Interacting Massive Particles (WIMPs)~\cite{Feng}. Thermally produced in the early Universe, these stable elementary particles should account for the relic density and consequently have a cross-section of the weak scale and a mass within a typical range of 10 GeV/c$^2$ to 1 TeV/c$^2$.  Liquid xenon experiments stand now as a leader in such high-mass WIMP searches thanks to both scalable highly radiopure absorbers to large masses and low background levels ensured by self shielding~\cite{xenon100-2016, Lux2016-2, PandaX-II-2016}. 
However, there is an increasing gain of interest for the search of low-mass WIMPs arising on the one hand from non evidence yet for supersymmetry at the LHC and on the other hand from new theoretical approaches favouring lighter candidates~\cite{Essig, Cheung, Hooper}. As an example, asymmetric dark matter models linking the relic density to the baryon asymmetry predict dark matter particles of a few GeV/c$^2$~\cite{Falkowski, Petraki, Zurek}. \\
A wide region of the parameter space $(\sigma_{SI},m_{W})$ giving spin-independent WIMP-nucleon cross-sections as a function of WIMP mass is thus yet to be explored at such low WIMP masses. A division of work is taking shape in the hunt for dark matter particles: an exploration of the high-mass region led by experiments with liquid scintillators, and light WIMP models to be tested by cryogenic detector experiments. Concerning the low- and intermediate-mass WIMP region between 0.8 and 20~GeV/c$^2$, the current situation is the one presented in
Fig.~\ref{fig:SotA2016}. It shows current experimental constraints: 90\%~C.L.  upper  limits~\cite{DAMIC, CRESST, SuperCDMS2014, CDMSlite2015, xenon100-2016, Lux2016-2, PandaX-II-2016, EDWlikelihood, Zeplin, DarkSide50} and closed contours~\cite{cdmsSI-contour, CoGeNT-contour, CRESST-contour, DAMA-contour} on the $(\sigma_{SI},m_{W})$ plane below 20 GeV/c$^2$, as well as the so-called neutrino floor~\cite{Billard}. Solid state detectors such as DAMIC~\cite{DAMIC} and particularly cryogenic such as the ones used by the CRESST~\cite{CRESST}, SuperCDMS~\cite{SuperCDMS2014, CDMSlite2015} and EDELWEISS~\cite{EDWlikelihood, EDWlowmass} experiments are potentially well suited to reach very low nuclear energy thresholds. These collaborations are actively working on the optimization of their experiments to focus on low mass WIMP searches, whereas the required thresholds seem to be more difficult to achieve for liquid Xe and Ar Time Projection Chambers (TPCs)~\cite{DarkSide50, xenon100-2016, Lux2016-2, PandaX-II-2016}, intrinsically limited by insufficient light scintillation efficiency. \\
\begin{figure*}[htb]
\includegraphics[width=1.\textwidth]{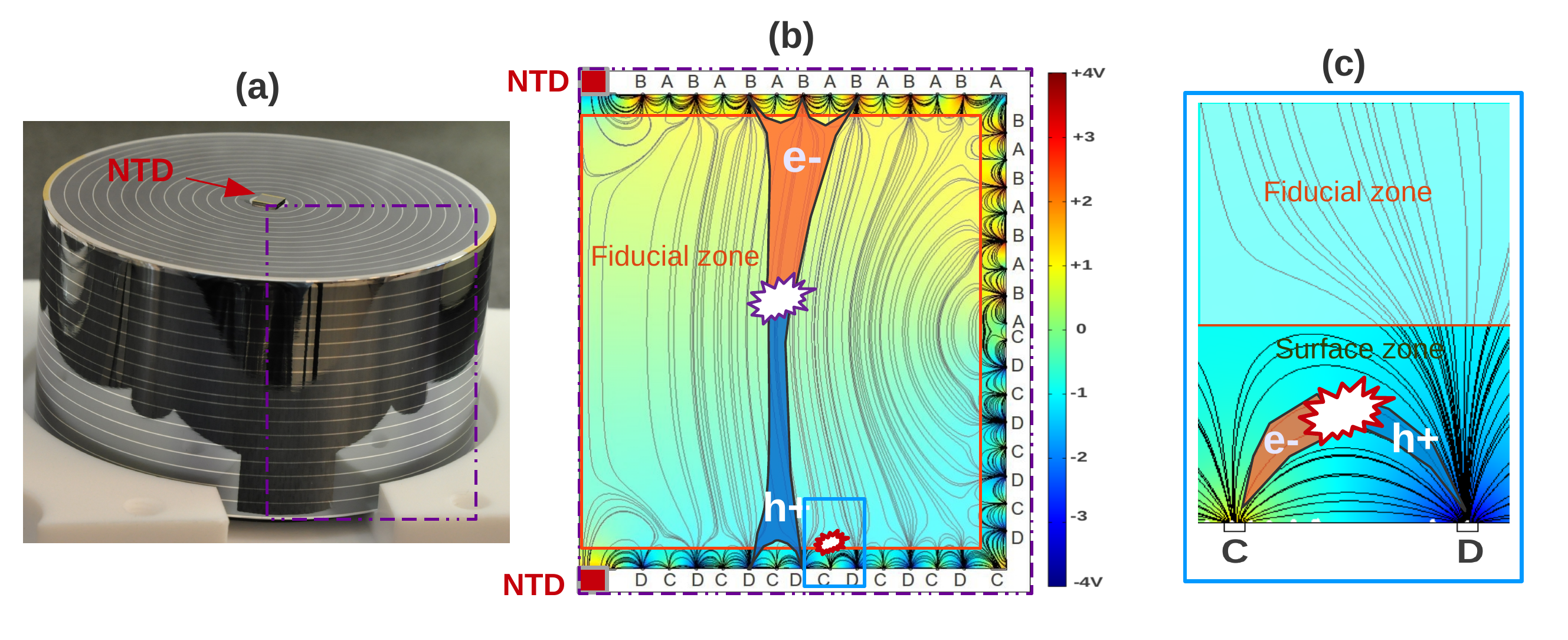}
\caption{\label{fig:fidscheme}Left panel (a): picture of an FID detector. Middle panel (b): cross-section of an FID detector as indicated by the dashed purple lines on panel (a). The zone in semi-transparency delimited by orange lines indicates the fiducial zone (i.e. where an energy deposit leads to charge collection on fiducial electrodes B and D as schematized on the picture). The color code gives the electric potential map. Right panel (c): illustration of charge collection for a surface event (i.e. involving at least one veto electrode).}
\end{figure*}
In this new context, the EDELWEISS experiment originally designed for the search of WIMPs of $\mathcal{O}(100\;\mathrm{GeV/c}^2)$ has undergone a redirection of its  strategy. The present paper thus proposes a roadmap dedicated to the optimization of EDELWEISS detectors for low-mass WIMP searches. To define R\&D priorities, projected sensitivities are computed spanning the detector performance achievable at short-term. This paper is organized as follows: in section~\ref{sec:edelweiss}, we first describe the EDELWEISS experiment and present the background, signal and detector response models used to simulate the ouputs of future data acquisition. We then present in sections~\ref{sec:bdt} and \ref{sec:likelihood} the two analysis methods used to derive projected sensitivities, namely the boosted decision tree (BDT) and the profile likelihood, respectively, and compare their performance in section~\ref{sub:analysisstrategy}. The study of EDELWEISS detector optimization is presented in section~\ref{sec:optimizing} where the impact on the sensitivity of detector performance is reviewed. Finally we present in section~\ref{sec:prospects} the expected sensitivity that can be achieved by 2018, and discuss the requirements for a larger scale experiment to probe the space parameter down to the bound imposed by the coherent solar neutrino scattering~\cite{Billard}, either in terms of background for WIMP search or as a potential neutrino study.

      \section{The EDELWEISS-III experiment}
\label{sec:edelweiss}
The EDELWEISS-III dark matter direct detection experiment~\cite{performance-paper} is installed in the deepest European underground laboratory, the Laboratoire Souterrain de Modane (LSM). It houses the largest operating mass of germanium detectors devoted to the search for dark matter with twenty-four $820-890$~g high purity Ge cylindrical crystals, each with a diameter of 7~cm and a height of 4~cm, called FID (Fully Inter-Digitized) detectors.
These are cooled down to cryogenic temperatures (18 mK) in order to perform a double measurement of ionization and heat signals, which is used to discriminate nuclear recoils induced by WIMP elastic scattering on Ge nuclei from electronic recoils induced by $\beta$- and $\gamma$-rays. Charge collection is carried out by concentric Al electrodes interleaved on all the absorber surfaces (see Fig.~\ref{fig:fidscheme}(a)). 
Electrodes are alternatively biased in such a way as to produce a field structure that defines two regions recognizable from the pair of electrodes involved in charge collection: \newline
- a fiducial zone where the created carriers drift towards the fiducial electrodes B and D, as shown in Fig.~\ref{fig:fidscheme}(b). \newline
- a surface zone where an energy deposit leads to a charge collection shared between one fiducial electrode and one so-called veto electrode: either (A\&B) or (C\&D), as shown in Fig.~\ref{fig:fidscheme}(c). \newline
The readout of the four types of electrodes allows fiducial selection of events and results in a background rejection factor for surface $\alpha$- and $\beta$-events of $4 \times 10^{-5}$ and $2.5 \times 10^{-6}$, respectively~\cite{performance-paper}.
FID detectors are also equipped with two NTD (Neutron Transmutation Doped) Ge sensors glued on their two planar surfaces, allowing to measure temperature elevations of a few $\mu \mathrm{K}$ that characterize energy deposits of the order of one keV. 
In addition to inducing the ionization signal, the drift of the $N_p$ electrons and holes created following a particule interaction amplifies the heat signal through Neganov-Luke effect~\cite{Luke}. Neglecting trapping effects~\cite{charge-paper}, the full conversion into thermalized phonons of the work done on the charge carriers during the drift produces an additional heat contribution $E_{Luke}$ equal to:
\begin{equation}
E_{Luke}=N_peV=Q(E_r)\frac{E_r}{\epsilon_{\gamma}}~eV
\end{equation}
where $e$ is the elementary charge, $V$ the collection bias and $Q(E_r)$ the ionization yield associated to the recoil energy $E_r$.  The quantity $\epsilon_{\gamma}$ = 3~eV per electron charge is the average energy required to create  an $e^-/h^+$ pair for electronic recoils in germanium~\cite{Knoll}.  
This quantity is approximately four time less than the energy required by a nuclear recoil to produce a pair, a factor that is taken into account by the normalized ionization yield factor $Q(E_r)$. 
The total energy of the heat signal is thus:
\begin{equation}
E_{heat} = E_r + E_{Luke} = E_r \times \left(1 + \frac{Q(E_r)~V}{3}\right)
\end{equation}
In the limit of biases up to 100~V, $E_{Luke}$ dominates and both phonon and ionization signals become proportional to $N_p$, effectively losing the discrimination power offered by the double measurement. To prevent this, the detectors are commonly operated at biases of a few volts.
However, in the context of low-mass WIMP searches, the optimal bias needs to be re-evaluated in view of the constraints imposed by the experimental backgrounds and the required thresholds. 
To answer this question, a modeling of both EDELWEISS-III backgrounds and signal, as well as the FID detector response, has to be carried out as described in subsections~\ref{sub:backgroundmodel}, \ref{sub:signalmodel} and \ref{sub:detmodel}, respectively.

       \subsection{Background model}
\label{sub:backgroundmodel}
\begin{figure}[htbp!]
\includegraphics[width=0.49\textwidth]{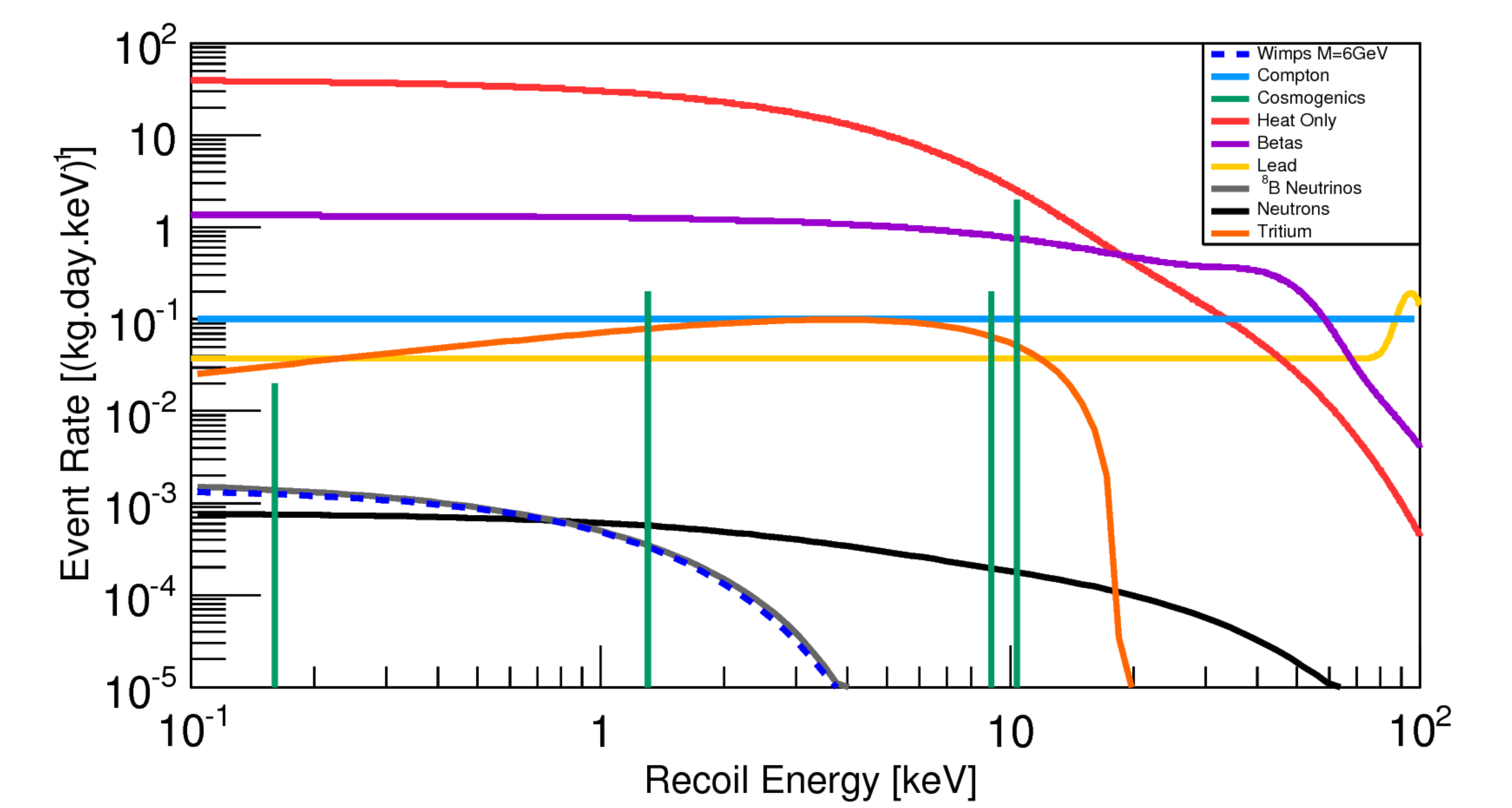}
\caption{\label{fig:recoilenergybackgrounds}Event rate for a total exposure of 1 $\mathrm{kg\cdot d}$ as a function of the recoil energy. Solid lines correspond to the recoil energy spectra of the different background components as indicated by the color code. The blue dashed-line shows the theoretical spectrum of a 6~GeV/c$^2$ WIMP with $\sigma_{SI}=4.4.\times 10^{-45}\;\mathrm{cm}^2$, which is extremely similar to the one of solar $^8$B neutrinos represented in gray solid-line.}
\end{figure}

Our background model is data driven by the first physics run of the EDELWEISS-III experiment~\cite{EDWlikelihood, EDWlowmass} and based on sidebands. Each background component~\cite{silviaLRT, alexjLTD} is characterized by a spectral shape, an event rate and an ionization yield $Q(E_r)$. The latter, which is normalized to 1 for $\gamma$-rays, is given for all backgrounds in Table~\ref{tab:rates}, together with the expected number of events in the recoil energy range $[0,~20]~\mathrm{keV}$ considering a total exposure of 1 $\mathrm{kg\cdot d}$. The associated recoil energy spectra are shown in solid lines in Fig.~\ref{fig:recoilenergybackgrounds}  for each individual background described below. Analytic functions used to describe these backgrounds can be found in the appendix.

The three backgrounds considered first have been studied extensively to understand tritium decays and cosmogenic activation in germanium detectors, as described in~\cite{tritium}. These are:
\begin{itemize}
\item Compton induced electronic recoils, which are well described by a flat spectrum in the region of interest.\smallskip

\item Tritium $\beta$-decays, inducing electronic recoils, for which the recoil energy spectrum is parametrized using Eq.~\ref{Eq-tritium}, with an endpoint at 18.6~keV. \smallskip

\item Cosmogenic activation induced X-ray peaks following the electron capture from the K-, L- or M-shells in Ge detectors and producing lines whose intensity is strongly dependent on the history of the detector~\cite{tritium}.
In these projections we assume constant arbitrary rates for the K, L and M peaks, with fixed values of 10 (100) for K/L (K/M) electron capture intensity ratios~\cite{Bahcall, Bambynek}. In the simulation, the K peak is resolved as the so-called 10~keV triplet (at 10.37, 9.66 and 8.98~keV),  while the L and M are considered as single peaks at 1.30 and 0.16~keV, respectively (these lines have no impact on our following sensitivity study thanks to the resolution)~\cite{ToI}.
\end{itemize}
Background associated with surface events has been studied during low-mass WIMP analyses of the EDELWEISS-III data using the eight detectors having the best performance: associated energy spectra have been directly measured for top and bottom sides of each detector independently, according to their ionization topologies, and then extrapolated down to lower energy~\cite{EDWlowmass, EDWlikelihood}. The corresponding ionization yields have also been measured and the values are given in Table~\ref{tab:rates}. The dispersions around these $Q(E_r)$ values are dominated by experimental resolutions at low energy. In the present publication, the projected sensitivities are obtained using surface backgrounds as derived detector by detector by fitting the resulted averaged spectra weighted by the exposure with analytic functions, which allow reproducing the data:
\begin{itemize}
\item The spectrum associated with surface $\beta$-decays from the $^{210}$Pb decay chain has been derived by fitting the averaged spectrum in data between 5 and 50~keV with a function given in Eq.~\ref{Eq-surface-210Pb}. \smallskip
\item Surface $^{206}$Pb recoils from the $\alpha$-decays of $^{210}$Po are assumed to be produced in equilibrium with the $^{210}$Pb decay chain. The recoil energy spectrum has been derived by fitting the averaged spectrum in data between 10 and 100~keV, leading to a gaussian distribution associated with a flat component, see Eq.~\ref{Eq-surface-206Pb}.
\end{itemize}
Three other backgrounds have been modeled namely heat-only events and nuclear recoils arising from either neutrons or $^8$B solar neutrinos:
\begin{itemize}
\item The dominant background in the EDELWEISS-III low energy data is due to heat-only events characterized by no ionization signal. These are not to be mistaken with noise events resulting from triggering on upper fluctuations of the heat baseline. Investigations are still ongoing to clearly identify the origin of these events and eliminate them. In~\cite{EDWlikelihood}, background from heat-only events in the region of interest (ROI) has been modeled using a kernel density estimator (KDE) function of the data in the sideband with negative ionization energy. The spectrum is parametrized by the sum of two exponential functions (see Eq.~\ref{Eq-HO}). It has been noted that the shape of the heat-only background does not vary with bias voltage therefore allowing us to properly model them at 8~V, where we can separate them from other background components and extrapolate them at higher voltages. \smallskip
\item Radiogenic neutrons can produce single scatter nuclear recoils with the same ionization yield as WIMPs and thus mimic their signal. One considers that neutrons (as $^8$B neutrinos, see below) induce nuclear recoils with ionization yield values gaussian-distributed around $Q(E_r)=0.16E_r^{0.18}$~\cite{performance-paper}. The neutron spectral shape is obtained from a fit on the EDELWEISS-III GEANT4 simulations in the nuclear recoil energy range from 2 to 20~keV~\cite{EDWlikelihood}. It consists of the sum of two exponentials as given by Eq.~\ref{Eq-radiogenic}. The absolute single rate is derived from the number of multiple nuclear recoils observed in WIMP search data between $[10,100]$~keV~\cite{performance-paper}, multiplied by the single/multiple ratio of 0.45 provided by the same GEANT4 simulations~\cite{EDWlikelihood}. \smallskip
\item Background from coherent neutrino-nucleus scattering induced by solar $^8$B neutrinos, whose spectral shape is given in~\cite{ruppin}, can produce single scatter nuclear recoils with the same ionization yield $Q(E_r)=0.16E_r^{0.18}$ as neutrons or WIMPs. 
Also their spectral shape is extremelly similar to the one of a 6~GeV/c$^2$ WIMP with $\sigma_{SI}=4.4.\times 10^{-45}\;\mathrm{cm}^2$~\cite{ruppin}. Neutrinos from other sources are not included as they would have no impact on the sensitivities computed in the [0.8, 20] GeV/c$^2$ WIMP mass range considered for this study.\smallskip
\end{itemize}
In the present study, spectral shapes are assumed to be perfectly known. With the versatility of the EDELWEISS detectors,
it has been so far relatively easy to accumulate relevant calibration and sideband data for the study of the different backgrounds.
As a result, the precision on the background models is steadily increasing.  Systematic uncertainties on rates are fixed to 16\% for solar $^8$B neutrinos, 50\% for neutrons and 30\% for tritium. Systematics of 10\% are associated to the rates of all the remaining individual backgrounds and arise from the combination of measured statistical uncertainties and extrapolation (in the ROI) of spectral uncertainties~\cite{EDWlikelihood}.
\begin{table}[h!]
\caption{\label{tab:rates}Background model parameters for EDELWEISS Ge detectors}
\begin{center}
\begin{tabular}{|c|c|c|c|}
  \hline
 Background  & Ionization  & Event rate  $(\mathrm{kg\cdot d})^{-1}$ \\ 
 	type	 & 	yield  $Q(E_r)$ & for $E_r \in$~[0,~20~keV]  \\ 
  \hline
	Compton & 1 & 2.00 \quad \\
   \hline
 	 Tritium & 1 & 0.990  \quad \\
  \hline
 	Cosmogenic & 1 &  2.42 \quad \\
   \hline
   Beta & 0.4 & 16.1  \quad \\
  \hline
   Lead & 0.08 &  0.740 \quad \\
 
  \hline
     Heat-only & 0 & 145.  \quad \\
  \hline
  Neutron & $0.16E_r^{0.18}$ &  $4.80 \times 10^{-3}$ \quad  \\
  \hline
   $^8$B neutrino & $0.16E_r^{0.18}$ &  $1.37 \times 10^{-3}$
\quad  \\
  \hline
\end{tabular}
\end{center}
\end{table}
                                                                                  
\subsection{Signal model}
\label{sub:signalmodel}
To compare our projected sensitivities to previous experimental limits, we will assume the Standard Halo Model described by a truncated Maxwell-Boltzmann WIMP velocity distribution which, translated in the Earth frame, is defined by:
\begin{equation}
f(\vec{v}) = \left\{
\begin{array}{rrr}
\rm & \frac{1}{N_{\rm esc}(2\pi\sigma^2_v)^{3/2}}\exp\left[-\frac{\left(\vec{v} + \vec{V}_{\rm lab}\right)^2}{2\sigma^2_v}\right]  \\	& \;\;\;\;\;\;\;\ \text{if $|\vec{v} + \vec{V}_{\rm lab}|<v_{\rm esc}$}  \\
& \\
\rm & 0  	\;\;\;\;\;\;\; 	\ \text{if $|\vec{v} + \vec{V}_{\rm lab}|\geq v_{\rm esc}$}
\end{array}\right.
\end{equation}
where $\vec{v}$  is the WIMP velocity, $\sigma_v$ is the WIMP velocity dispersion related to the local circular velocity $v_0$ such that $\sigma_v = v_0/\sqrt{2}$, $\vec{V}_{\rm lab}$ and $\vec{v}_{\rm esc}$ are the laboratory and the escape velocities with respect to the galactic rest frame, and $N_{\rm esc}$ is the correction to the normalization of the velocity distribution due to the velocity cutoff ($v_{\rm esc}$). \\
The differential recoil energy rate is then given by~\cite{lewin}:
\begin{equation}
\frac{dR}{dE_r} = M T\times \frac{\rho_0 \sigma_0}{2m_{\chi}m^2_r}F^2(E_r)\int_{v_{\rm min}}\frac{f(\vec{v})}{v}d^3v
\end{equation} 
where $\rho_0$ is the local dark matter density, $m_{\chi}$ is the WIMP mass, $m_r = m_\chi m_N/(m_\chi + m_N)$ is the WIMP-nucleus reduced mass and $\sigma_0$ is the normalized nucleus spin-independent cross section. $F(E_r)$ is the nuclear form factor that describes the loss of coherence for recoil energies above $\sim$10~keV. In the following, we will consider the standard Helm form factor~\cite{lewin}. For the sake of comparison with running experiments, we will consider the standard values of the different astrophysical parameters: $\rho_0 = 0.3 $~GeV/c$^2$/cm$^3$, $v_0 = 220$~km/s, $V_{\rm lab} = 232$~km/s and $v_{\rm esc} = 544$~km/s~\cite{lewin,astroparam}.

        \subsection{Detector response model}
\label{sub:detmodel}
We build a simplified detector response model based on the capability of FID detectors to reject surface events using sets of interleaved fiducial and veto electrodes. We make no distinction between top and bottom surfaces and consider only two ionization measurements: the fiducial ionization energy $E_{fid}$  defined as the average of the energies measured on B and D channels and the veto ionization energy $E_{veto}$ refering indifferently to the energy measurement on the veto A or C involved in charge collection in case of a surface event. Both ionization and veto energy channels are characterized by baseline energy resolutions of $\sigma_{E_{fid}}$ and $\sigma_{E_{veto}}$, respectively, where $\sigma_{E_{veto}} = \sqrt{2}~\sigma_{E_{fid}}$. Ionization energies are expressed in $\mathrm{ keV_{ee}}$ (electron equivalent) as follows : 
\begin{equation}
E_{fid}=\alpha \,Q(E_r) \times E_r\quad \quad E_{veto}=\beta \,Q(E_r) \times E_r
\end{equation}
where $(\alpha,\beta)$ is $(1,0)$ for fiducial events and $(\frac{1}{2},1)$ for surface events. This model neglects the small fraction of events 
where $(\alpha,\beta)$ have slightly different values due to additional charge sharing between veto and fiducial electrodes from the ionization cloud in the crystal and from multiple scaterring that could be induced by neutrons and gammas. 

The heat energy measurement $E_{heat}$ is characterized by its baseline energy resolution $\sigma_{E_{heat}}$. It is useful for the analysis purpose to define a normalized heat energy $\tilde{E}_{heat}$ expressed as follows:
\begin{equation}
\tilde{E}_{heat}= E_r \times \left(1 + \frac{Q(E_r)~V}{3}\right) /\left({1+\frac{V_{fid}}{3}}\right)  
\label{Eq-Eheat}
\end{equation}
where V is the voltage difference between the two collecting electrodes ($V=V_{fid}$ for fiducial events and $V=V_{surf}$ for surface events). The electron-equivalent heat energy $\tilde{E}_{heat}$ is expressed in $\mathrm{ keV_{ee}}$ as it is normalized such that $\tilde{E}_{heat}=E_{fid}=E_r$ for electronic recoils in the fiducial volume.  \\
\begin{figure}[htbp!]
\includegraphics[width=0.51\textwidth]{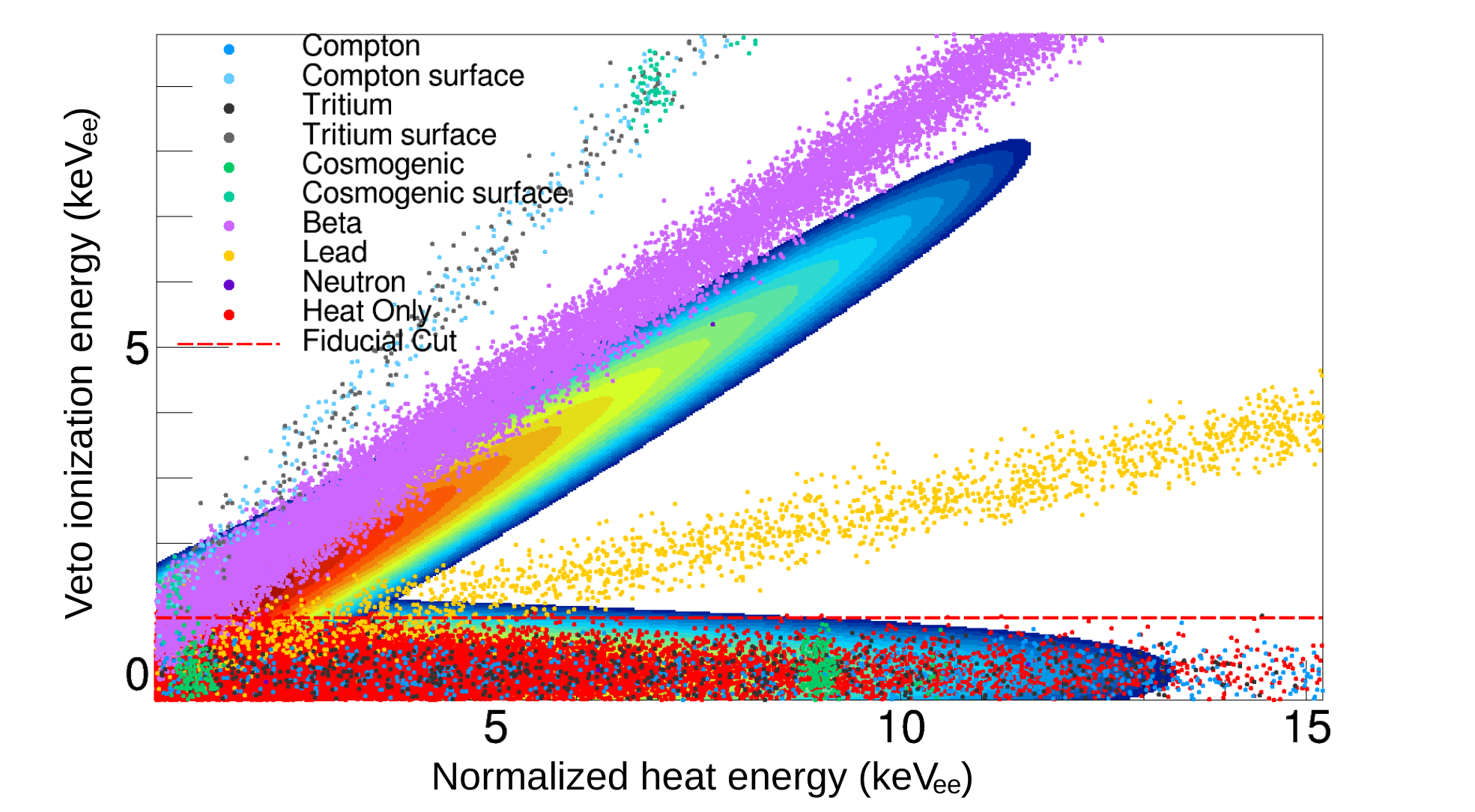}
\includegraphics[width=0.51\textwidth]{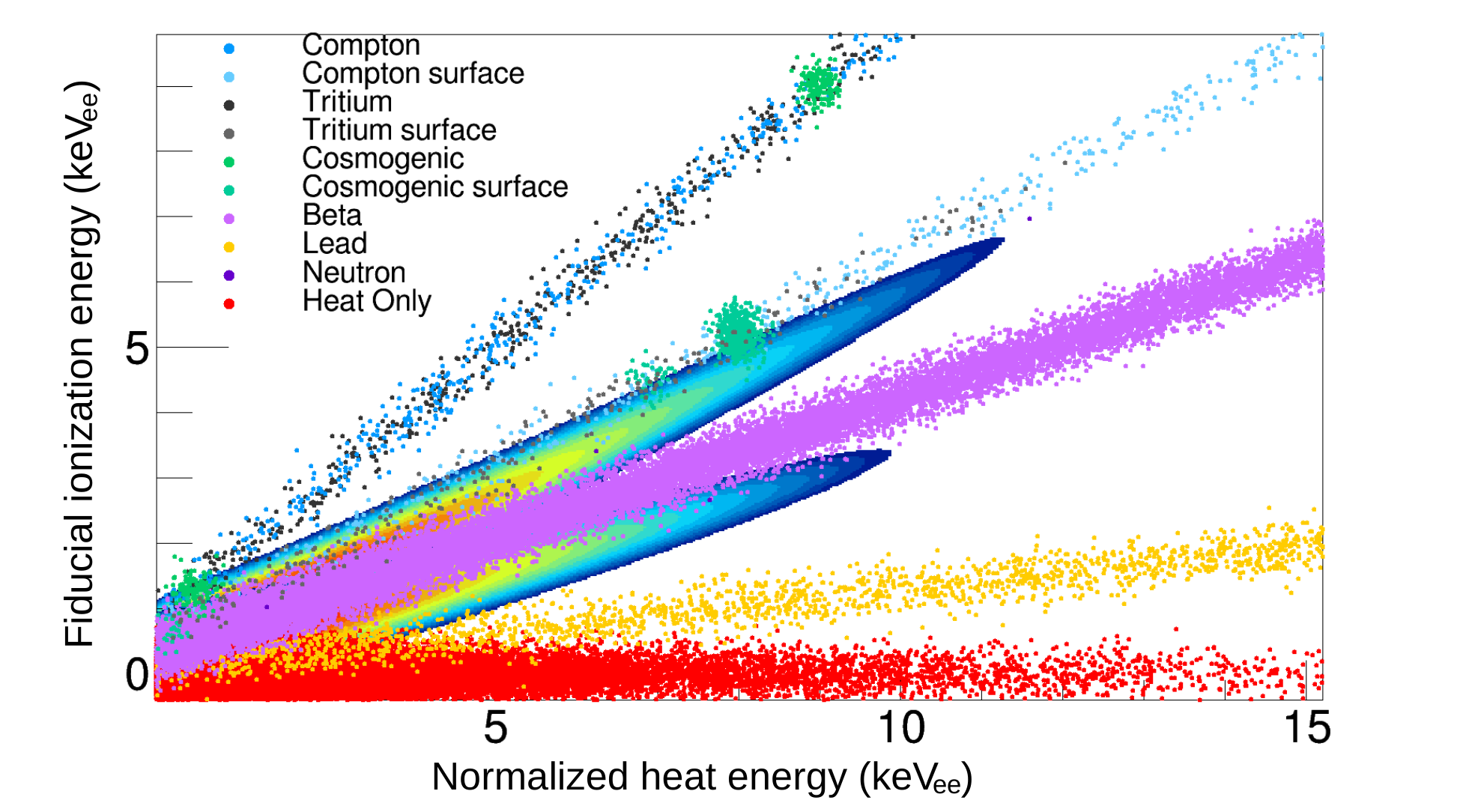}
\includegraphics[width=0.51\textwidth]{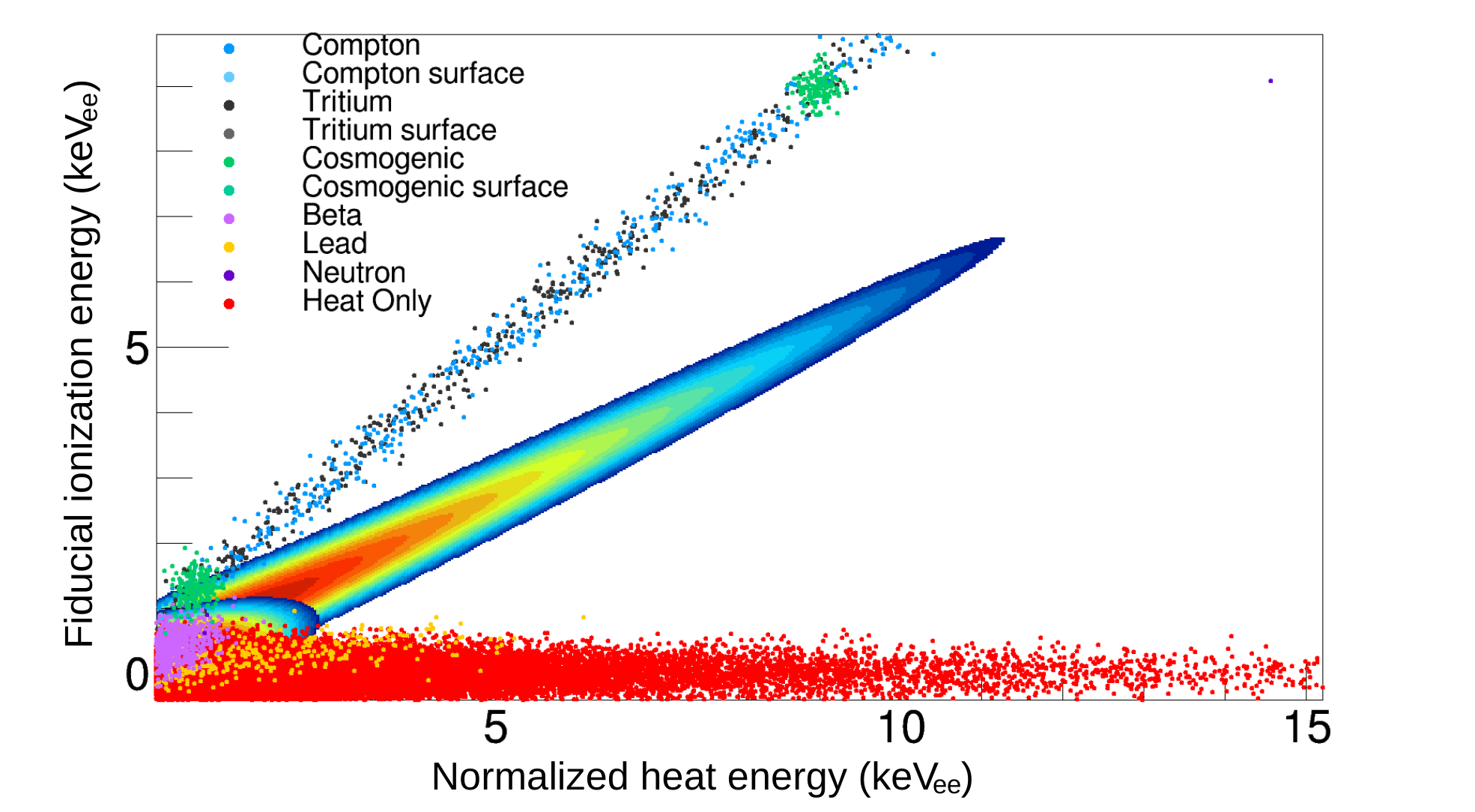}
\caption{\label{fig:BiPlot}Top panel: simulated data in the veto ionization energy ($E_{veto}$) vs. normalized heat energy ($\tilde{E}_{heat}$) plane before fiducial selection for a total exposure of 1000 $\mathrm{kg\cdot d}$ with standard FID performance: $\sigma_{E_{heat}} = 500$~eV, $\sigma_{E_{fid}} = 200~\mathrm{eV_{ee}}$, and a bias voltage   $V_{fid}= 8$~V. The fiducial cut is indicated by the red dashed line. Colored contours correspond to a theoretical 15 GeV/c$^2$ WIMP signal (both bulk and surface events), which is used as an illustration. Middle and bottom panels: same as top panel but in the fiducial ionization energy ($E_{fid}$) vs. normalized heat energy ($\tilde{E}_{heat}$) plane, before (middle) and after (bottom) fiducial selection.}
\end{figure}
Electronic and nuclear recoil discrimination is provided by the double measurement of $E_{fid}$ and $E_{heat}$. Surface event rejection is performed by requiring $E_{veto}<3\;\sigma_{E_{veto}}$. The heat energy lower bound corresponds to a 50\% threshold efficiency, assuming a trigger at $6\;\sigma_{E_{heat}}$. This strong requirement allows to ensure that events arising from noise fluctuations can be neglected.\\
In the following, different detector configuration scenarios will be considered with $V_{fid}$ as the detector bias voltage. The ratio $V_{fid}/V_{surf}=8/5.5$ is fixed such that the fiducial volume, determined by the electric field, corresponds to 75\% of the total crystal volume independently of the bias voltage~\cite{performance-paper}. 
Fig.~\ref{fig:BiPlot} presents the event distribution either in the veto ionization energy ($E_{veto}$) vs. normalized heat energy ($\tilde{E}_{heat}$) plane (top panel) or in the fiducial ionization energy ($E_{fid}$) vs. normalized heat energy ($\tilde{E}_{heat}$) plane (middle and bottom panels), for both the background model and a 15 GeV/c$^2$ WIMP (chosen as an illustration) using standard FID baseline resolutions and bias voltage. Middle and bottom panels of the figure correspond to before and after fiducial selection, respectively. In top panel of Fig.~\ref{fig:BiPlot}, the fiducial cut is represented by the red dashed line. One clearly sees that this cut is very efficient at cleaning the signal region above $\tilde{E}_{heat} = 2~\mathrm{keV_{ee}}$. However better performance, as improved energy resolutions, combined with sophisticated analysis methods will be necessary to probe the lowest energies, and therefore the low WIMP mass scenarios, as the signal and backgrounds start to overlap.
       \section{Analysis methods}
\label{sec:methods}
As dark matter experiments become increasingly sensitive to smaller and smaller WIMP-nucleon interaction cross-sections, requirements such as detector efficiency, discriminating variables and background levels are now extending to performant analysis methods. \\
The most common analysis approach in direct detection consists in defining a ROI in the parameter space separated from backgrounds and to consider any event recorded in this signal area as a WIMP candidate. In case no event is observed, an upper limit on the cross-section is reported with a 90\% confidence level (C.L.) corresponding from Poisson statistics to 2.3 WIMP events excluded. For a given signal acceptance, it gives the best limit an experiment can intend to achieve. However, experiments frequently report few events observed often compatible with possible background contaminations or noise fluctuations. In such cases, maximum-gap or optimum-interval methods formulated by S. Yellin~\cite{YellinOptimum} are preferentially used to optimize the exclusion limit. The advantage of these methods is that no assumption on backgrounds is required as only information relative to the expected signal spectrum and energies of WIMP candidates is used. However more competitive sensitivities can be achieved via multivariate analyses or statistical approaches, depending on the degree of knowledge of backgrounds. \\
Thus, to properly assess the potential of the EDELWEISS-III experiment, exclusion sensitivities will be derived from both boosted decision tree (BDT) and profile likelihood ratio approaches. These two analysis methods, described in sections~\ref{sec:bdt} and \ref{sec:likelihood}, respectively, can be considered as a pessimistic scenario and an optimistic one: in contrast to the BDT, the likelihood approach allows for statistical background subtraction assuming a perfect knowledge of each individual background spectral shape.
      \subsection{Boosted decision tree (BDT)}
\label{sec:bdt}
Boosted decision tree analysis belongs to machine learning techniques and is widely used to treat data in high energy physics (e.g.~\cite{Roe, Abazov, Conrad}). It is an extension of the commonly used cut-based selection strategy into a multivariate technique. Decision Tree analysis can be seen as a data classifier and is often used for signal/background discrimination. Indeed, as most events do not have all characteristics of either signal or background, the principle of Decision Tree is to keep events that fail a given criterion and check for other observable discriminants. Trees can then be boosted to combine weak classifiers into a new one with smaller error rate~\cite{Coadou}.
The result of a BDT analysis is given by a forest of $\text{N}_\text{tree}$ decision trees ($T_k$) combined into a unique output BDT score which varies from -1 (background-like) to +1 (signal-like) as presented in Fig.~\ref{fig:BDT-score} described in section~\ref{subsec:prospects-EDWbore8}. It reads as :
\begin{equation}
\text{BDT output} = \sum_{k=1}^{\text{N}_\text{tree}}\alpha_kT_k(\vec{E})
\label{BDT_output}
\end{equation}
where $\vec{E}$ refers to the set of considered observables associated to each recorded events $\{E_{fid}, E_{veto}, \tilde{E}_{heat} \}$ and $\alpha_k$ corresponds to the weight given to the $T_k$ classifier based on misclassification rate. In a procedure called boosting, misclassified events in a tree are given a higher weight before growing the next tree, in order 
to reduce the misclassification rate. Note that this rate can be reduced down to zero, but at the price of overtraining. This appears when trees contain too many leafs with a low number of events. In this case, the algorithm is sensitive to statiscal fluctuations of the training sample. A Kolmogorov-Smirnov (KS) statistical test~\cite{KStest} on cumulative distributions associated to the test sample and to the training sample is used to check for overtraining. In the following, even though we checked that the KS $p$-values were always greater than the two-sigma level, all the results given below are produced only from the test samples.\\
Thanks to its high level of reliability, ease of use use via the TMVA software~\cite{TMVA}, and robustness against mis-modeling of the background, BDT has recently started to be used in direct dark matter searches~\cite{Agnese:2014aze, EDWlowmass}. As described above, its robustness against background mis-modeling comes from the fact that even though the resulting ROI is tuned according to both the background and signal models but as no background subtraction procedure is applied, any mis-modeling of the background would result in a non-optimized, and therefore weaker, exclusion limit. \\
To compute EDELWEISS-III expected sensitivities, the BDT has been specifically trained for each WIMP mass and experimental condition using $10^6$ events generated by Monte Carlo in the 3D-space ($E_{fid},E_{veto},\tilde{E}_{heat}$) according to signal and background models. The fiducial cuts as described in section~\ref{sub:detmodel} are not applied. Instead we let the BDT learn by itself how to optimize the use of the three input variables to maximize the sensitivity to WIMPs. The exclusion limit can then be obtained from Poisson counting statistics by tuning the only remaining cut on the BDT score following:
\begin{equation}
\mu_{exc}(\text{cut}) = \frac{\sum_{n=0}^\infty\mu_{90}(n)\times P\left[\mu_B^{tot}(\text{cut}))| n\right]}{\epsilon_{\rm WIMP}(\text{cut})}
\end{equation}
where $\mu_{exc}$ refers to the excluded number of WIMP events at the 90 \% C.L. as a function of the BDT score cut, $\mu_{90}$ corresponds to the Poisson upper limit at the 90 \% C.L. derived for $n$ observed events, $P$ is the Poisson probability of observing $n$ events from $\mu_B^{tot}$ expected background events (also as a function of the BDT score cut), and $\epsilon_{\rm WIMP}$ is the WIMP signal efficiency which decreases from 1 to 0 when varying the BDT score cut from -1 to +1. Finally, we integrate over all possible outcomes of observed events by summing over $n$ from 0 to infinity. Therefore, $\mu_{exc}(\text{cut})$ as a function of the BDT score cut naturally exhibits an optimal point, where it is minimal, from which we derive the optimal BDT score cut to be used in the limit calculation. This procedure has been first intoduced in~\cite{Agnese:2014aze}. 
       \subsection{Maximum likelihood analysis}
\label{sec:likelihood}
We now consider a maximum likelihood frequentist approach known as profile likelihood ratio. Toy data are generated from Monte Carlo simulations according to both background and detector response models even though, contrary to the BDT, only events passing the fiducial cut are selected (see section~\ref{sub:detmodel}). The likelihood function and the test statistic used are described in the sections below.
\subsubsection{Unbinned likelihood function}
\label{subsubsec:likelihoodfunction}
An unbinned likelihood function is used to extract the whole accessible spectral information from the two registered energies $\{\tilde{E}_{heat},\ E_{fid}\}$ for each simulated event. Therefore, for a given cross-section $\sigma$, WIMP mass and exposure, the extended likelihood function is written as:
\begin{eqnarray}
    \mathcal{L}(\sigma,\vec{\mu}_B)  &=&     \mathrm{exp}({-(\mu_{S}+\sum_{j=1}^{M}\mu_{B}^j)}  ) \nonumber \\
    & \times &\prod_{i=1}^{N} \left(  \mu_S f_{S}(\vec{{E}_i}) +\sum_{j=1}^{M}\mu_{B}^j f_{B}^j(\vec{{E}_i}  ) \right)\nonumber \\
    &\times &    \prod_{j=1}^{M} \frac{1}{\sqrt{2\pi}\sigma_{\tilde{\mu}_B}^j} \mathrm{exp}\left\lbrace -\frac{1}{2}{\left(\frac{\mu_B^j-\tilde{\mu}_B^j}{{\sigma}_{\tilde{\mu}_B}^{j}}\right)}^2\right\rbrace \label{eq:likelihoodfunction}
  \end{eqnarray}
where $\mu_S$ and $\mu_B=\sum_j\mu_{B}^j$ correspond to the expected number of WIMP signal and background events from the model, respectively, with each background component labelled $j$, and are thus ajustable parameters with $M$ the number of background components used in the model. $\vec{E_i}$ refers to the set of observables $\{\tilde{E}_{heat,i}, E_{fid,i} \}$, depending on the readout and analysis considered, associated to each of the $N$ simulated nuclear recoils. The two first terms of Eq.~\ref{eq:likelihoodfunction} account for both Poisson fluctuations on the total number of observed events and spectral shape information through the probability density functions $f_{S}$ of WIMPs and $f_{B}^j$ of each background component. The last term of Eq.~\ref{eq:likelihoodfunction} allows to constrain the model parameters by including the knowledge of the different background rates with $\tilde{\mu}_B^j$ and ${\sigma}_{\tilde{\mu}_B}^{j}$ being respectively the expected number of background events and its associated systematic uncertainty. 
         \subsubsection{Likelihood ratio test statistic}
\label{subsub:teststatistic}
Following the statistical procedure described in~\cite{Likelihood}, $H_{\sigma}$ refers to the signal hypothesis where the WIMP-nucleon cross-section $\sigma$ can be non-zero and $H_{0}$ is the alternative null-hypothesis where $\sigma=0$. To test the compability between $H_{\sigma}$ and the best-fit model to the data, a hypothesis test is used, which is based on the profile likelihood ratio defined in Eq.~\ref{eq:hyptest}:
\begin{equation}
\mathrm{\lambda(\sigma)}=\frac{\mathcal{L}( \sigma,\Hat{\Hat{\theta}} )}{\mathcal{L}( \Hat{\sigma},\Hat{\theta} )}
\label{eq:hyptest}
\end{equation}
where $\theta$ represents the set of nuisance parameters which in our case refers to the expected number of background events from the model ($\theta  \equiv \{ \mu_B^j\}$). $\Hat{\theta}$ and  $\Hat{\sigma}$ are the maximum likelihood estimators of our nuisance and interest parameters, respectively. $\Hat{\Hat{\theta}}$ denotes the values of $\theta$ that maximize the conditional likelihood function for the specified cross-section value $\sigma$, i.e. we are profiling over the nuisance parameters. The test statistic $q_\sigma$ is then defined as:
\begin{equation}
q_\sigma=\left\{
  \begin{array}{ccc}
    \mathrm{-2ln(\lambda(\sigma))} & & \Hat{\sigma} \leq \sigma \\
    0 & &    \Hat{\sigma}> \sigma\\
  \end{array}
\right.
\end{equation} 
As one can deduce from this test, a large value of the test statistic $q_{\sigma}$ implies a large inconsistency between data and the tested hypothesis $H_{\sigma}$ such that the larger is the value of the test statistic $q_{\sigma}$, the higher is the confidence level at which the tested cross section $\sigma$ is excluded. The confidence level at which the tested cross-section $\sigma$ is excluded is given by $\alpha \%=1-p_s$ where $p_s$ is the signal $p$-value defined as follows:
\begin{equation}
p_s=\int_{q_{\sigma}^{obs}}^{\infty}f(q_{\sigma}|H_{\sigma})\mathrm{d}q_{\sigma}
\end{equation}
where $f(q_{\sigma}|H_{\sigma})$ is the probability density function of $q_{\sigma}$ under the hypothesis $H_{\sigma}$.  According to Wilk's theorem, $f(q_{\sigma}|H_{\sigma})$ asymptotically follows a half $\chi^2$ distribution with one degree of freedom, as described by Eq.~\ref{eq:chi2}, in the limit of large enough statistics.
\begin{equation}
f(q_\sigma|H_{\sigma})=\frac{1}{2} \delta(q_{\sigma})+\frac{1}{2}\frac{1}{\sqrt{2\pi}}\frac{1}{\sqrt{q_{\sigma}}}e^{-\dfrac{q_{\sigma}}{2}}
\label{eq:chi2}
\end{equation} 
We have checked from Monte Carlo simulations that this asymptotic approximation is well recovered even in the case where only few WIMP events are excluded. Since the distribution $f(q_{\sigma}|H_{\sigma})$ is known, Monte-Carlo simulations are only performed under the background-only hypothesis $H_0$ to determine $q_{\sigma}^{obs}$. The hypothesis $H_{\sigma}$ is rejected at $90\%~\mathrm{C.L.}$ if $p_s\leq10\% \leftrightarrow q_{\sigma}^{obs}\geq1.64$. The procedure described hereabove is repeated 500 times for each WIMP mass and considered scenarios. From the obtained set of cross-sections $\{\sigma_{90}\}$ excluded at $90\%~\mathrm{C.L.}$, the expected sensitivity is determined as $\sigma_{excl}$=median($\{\sigma_{90}\}$). 
%
         \subsection{Comparison \& strategy}
\label{sub:analysisstrategy}
Considering achievable detector performance according to ongoing R\&D carried out by the EDELWEISS collaboration presented in section~\ref{subsec:prospects-EDW2018}, sensitivity curves have been derived from both BDT and likelihood methods.  \\
\begin{figure}[htb]
\includegraphics[width=0.52\textwidth]{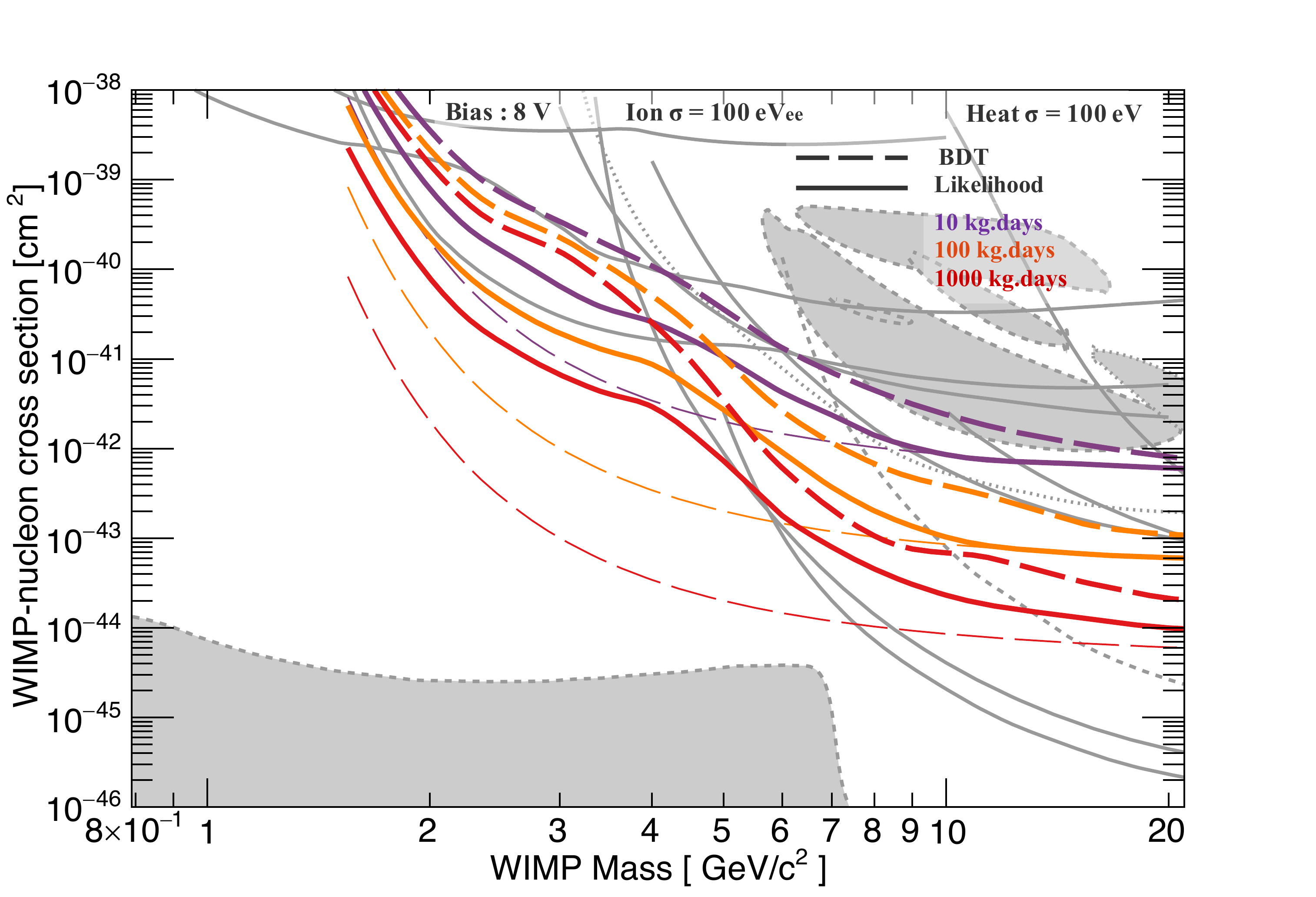}
\caption{\label{fig:LikeBDT}Comparison of the progression on exclusion limits with the exposure derived either from likelihood (thick solid lines) or BDT (thick dashed lines) analyses. Three values of the exposure are considered: 10~$\mathrm{kg\cdot d}$ (purple), 100~$\mathrm{kg\cdot d}$ (orange) and 1000~$\mathrm{kg\cdot d}$ (red). Concerning detector configuration, we use standard bias ($V_{fid}=8\;\mathrm{V}$) but expected energy resolutions ($\sigma_{E_{heat}}=100\;\mathrm{eV}$, $\sigma_{E_{fid}}=100\;\mathrm{eV_{ee}}$). The thin dashed lines correspond to the background-free sensitivity. All upper limits, contours and regions plotted in gray in this figure are those described in Fig.~\ref{fig:SotA2016}.}
\end{figure}
Fig.~\ref{fig:LikeBDT} presents a detailed comparison between the sensitivity curves derived with standard bias value of $V_{fid}=8\;\mathrm{V}$ from the likelihood (thick solid lines) and the BDT (thick dashed lines) analyses  considering achievable energy resolutions ($\sigma_{E_{heat}}=100\;\mathrm{eV}$, $\sigma_{E_{fid}}=100\;\mathrm{eV_{ee}}$, see section~\ref{subsec:prospects-EDW2018}) and varying the exposure from 10 $\mathrm{kg\cdot d}$ to 1000 $\mathrm{kg\cdot d}$ as indicated by the color code. Background-free sensitivites are shown in thin dashed lines. As expected, the likelihood analysis provides much more stringent limits at all considered WIMP masses.
The relative sensitivity gain varies with the exposure up to more than one order of magnitude between the extreme considered exposures such that, with current EDELWEISS-III backgrounds, a likelihood analysis with only 10 $\mathrm{kg\cdot d}$ is preferable to a BDT analysis with 1000 $\mathrm{kg\cdot d}$ for a WIMP mass below 4 GeV/c$^2$. This is due to a saturation effect of the exclusion limits derived from the BDT that indicates the presence of limiting backgrounds. Increasing the exposure does not lead to further improvement of the sensitivity to low-mass WIMPs as the ROI starts to be totally overwhelmed by background events. 
Exclusion limits obtained via the likelihood analysis, though affected by backgrounds, do not suffer from this saturation effect and allow to progress by one order of magnitude when the exposure is increased by a factor 100. This typical progression with the square root of the increase in exposure is allowed by the statistical subtraction of backgrounds with the likelihood whereas the BDT cut efficiency decreases rapidly as the cut becomes increasingly stringent. 
At higher mass, the relative sensitivity gain conferred by the likelihood approach is still noticeable but less marked, indicating that it is possible to explore WIMP masses above 4-5 GeV/c$^2$ with EDELWEISS even if the knowledge of the backgrounds does not reach the precision needed for performing a likelihood analysis.  \\
Note that even the exclusion limits derived from the likelihood analysis ineluctably saturate at some point, however at higher exposure than when derived from a BDT approach. This saturation effect appears in presence of a background similar in shape to the signal (as background from $^8$B solar neutrinos), which disables the spectral discrimination. \\
Looking at the efficiency of both analysis methods, it appears clearly that in order to assess the full potential of the EDELWEISS-III experiment, the likelihood analysis is to be privileged to span the various experimental conditions.
       \section{Optimizing the sensitivity to low-mass WIMPs}
\label{sec:optimizing}
According to the conclusions of section~\ref{sub:analysisstrategy}, the superiority of the likelihood method is a general result in all the performed studies, and for brevity, we will omit the BDT plots in the next subsections  
where the effects of various experimental factors on the sensitivity to WIMP masses below 20 GeV/c$^2$ are studied.  
In subsection~\ref{sub:boosting}, we describe the impact of the threshold on the sensitivity, in particular, the effect of reducing it through Neganov-Luke boosting. We then briefly review, respectively in subsections~\ref{sub:backgroundimpact} and \ref{sub:resolutions}, how each individual background affects the sensitivity and the expectations from energy resolution improvements. Finally, we explore in subsection~\ref{sub:readout} the actual benefits of both the ionization and heat double measurement and the surface rejection capability, to determine in which mass range the FID detector design is required for low-mass WIMP searches.
       \subsection{Thresholds and Neganov-Luke boost}
\label{sub:boosting}
Reducing thresholds is a common objective shared by all dark matter experiments as the theoretical recoil energy spectrum falls typically with an exponential behaviour. It is compulsory for low-mass WIMP searches as the spectrum is increasingly softer as the WIMP mass gets lower. The Neganov-Luke boost can be used to lower thresholds by amplifying the signal through the application of high voltage biases on collecting electrodes. For Ge or Si detectors, the amplification gain provided by the increase of the collection-bias between two electric potentials  $V_1$ and $V_2$  is $(1+Q(E_r)\, V_2/\epsilon_{\gamma}) / (1+Q(E_r)\, V_1/\epsilon_{\gamma})$. However, since the Neganov-Luke effect linearly depends on the number of charge carriers, its enhancement tends to transform the heat measurement into a pale copy of the ionization measurement and thus gradually disables the discrimination between nuclear and electronic recoils.

\begin{figure}[htb]
\includegraphics[width=0.52\textwidth]{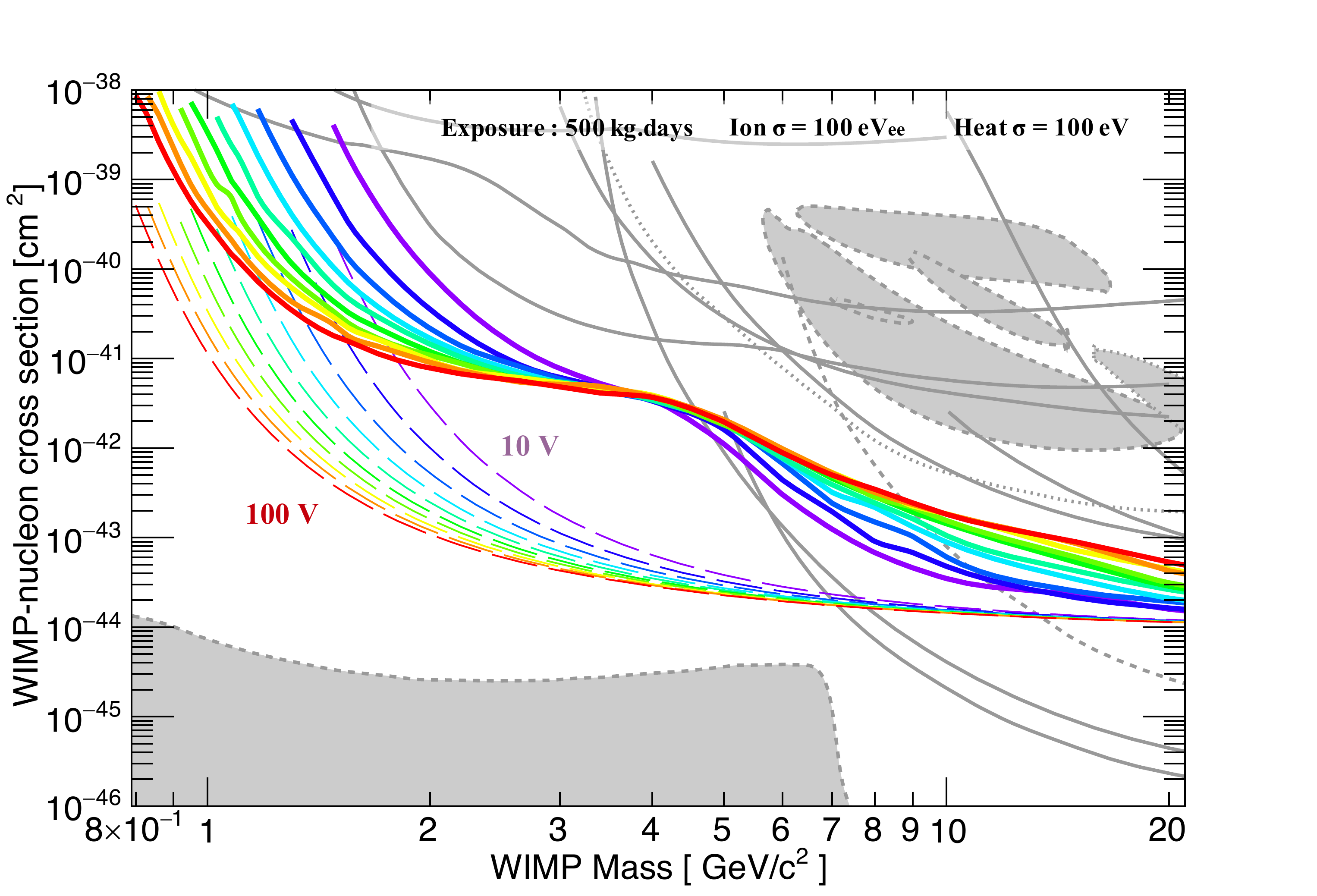} 
\caption{\label{fig:MultiVolt}Influence of the bias voltage on the sensitivity at fixed ionization and heat resolutions ($\sigma_{E_{heat}}=100\;\mathrm{eV}$, $\sigma_{E_{fid}}=100\;\mathrm{eV_{ee}}$). The color code indicates the bias condition rising with a 10~V step from 10~V in purple to 100~V in red. Solid and dashed lines refer to the exclusion limits and to the background-free sensitivities for an exposure of 500 $\mathrm{kg\cdot d}$. All upper limits, contours and regions plotted in gray in this figure are those described in Fig.~\ref{fig:SotA2016}.}
\end{figure}
\begin{figure*}[h]
\includegraphics[width=0.49\textwidth]{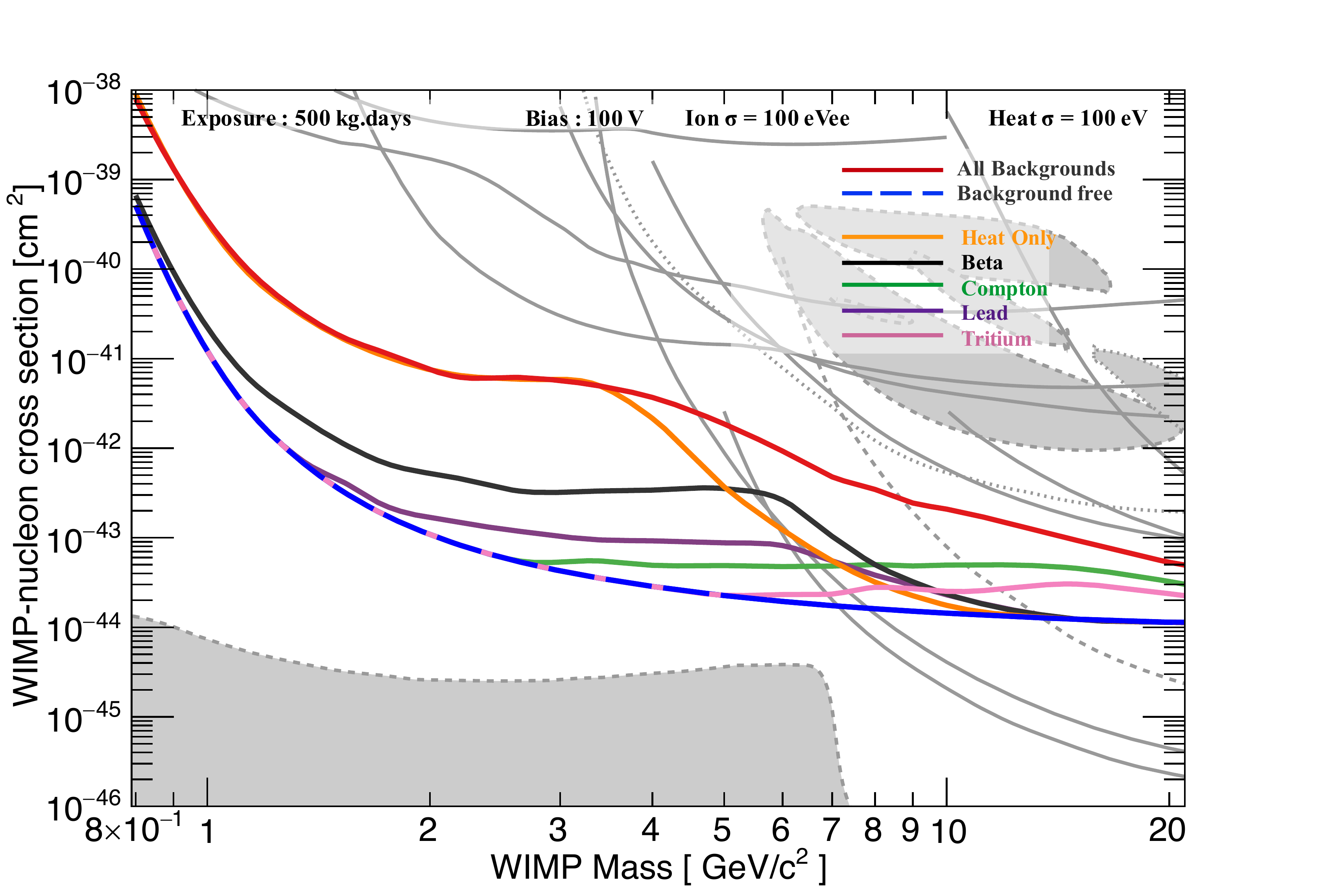}
\includegraphics[width=0.49\textwidth]{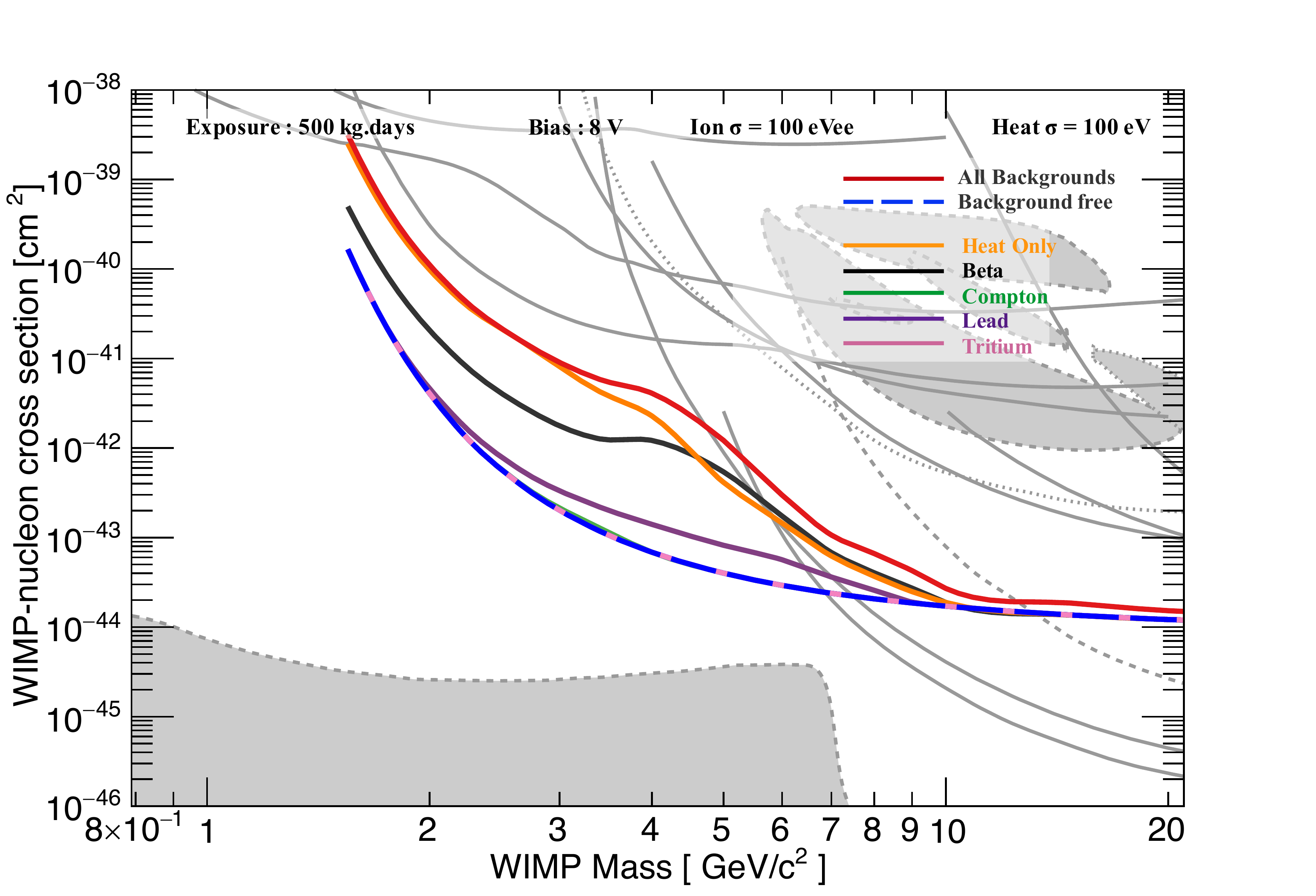} 
\includegraphics[width=0.49\textwidth]{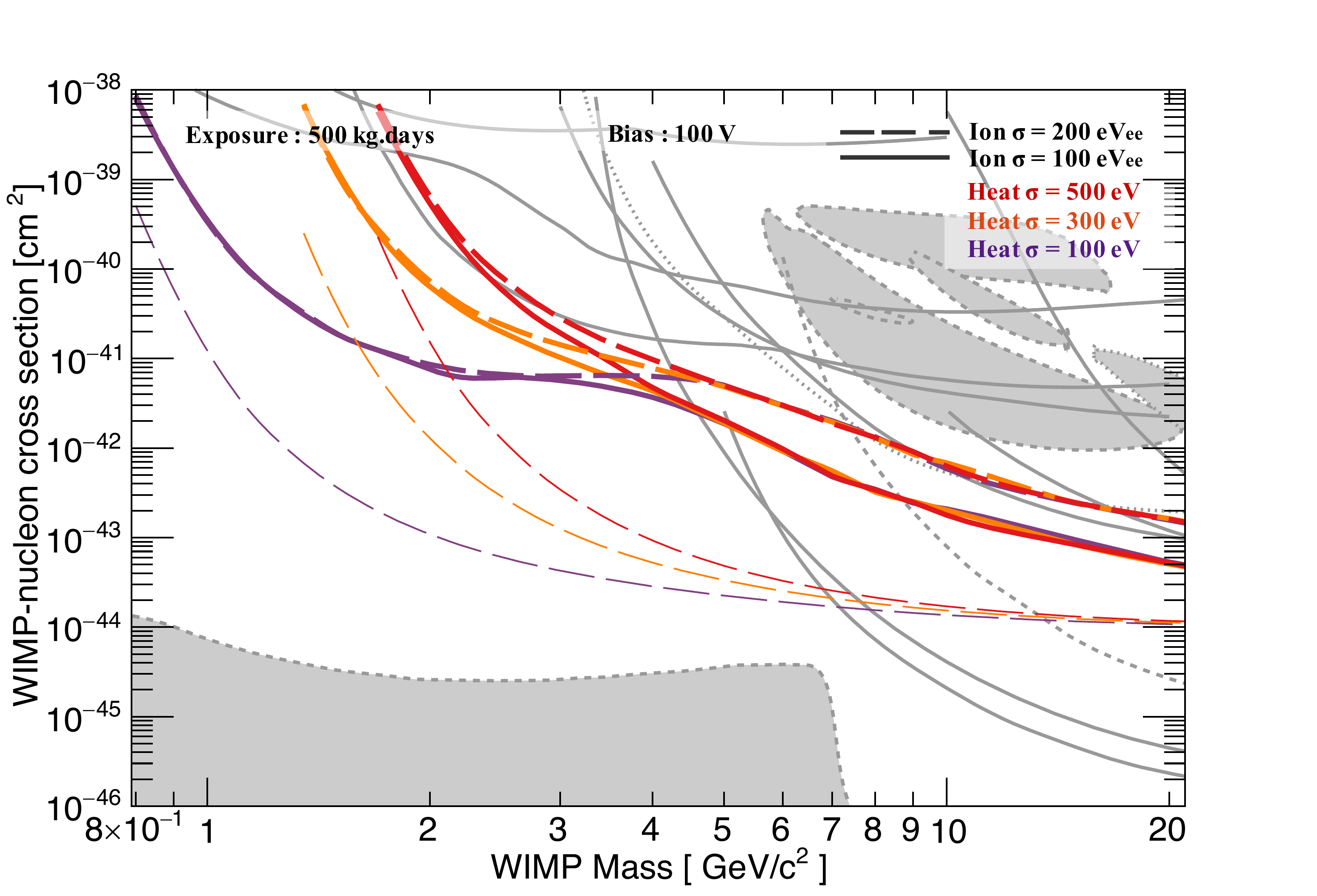} 
\includegraphics[width=0.49\textwidth]{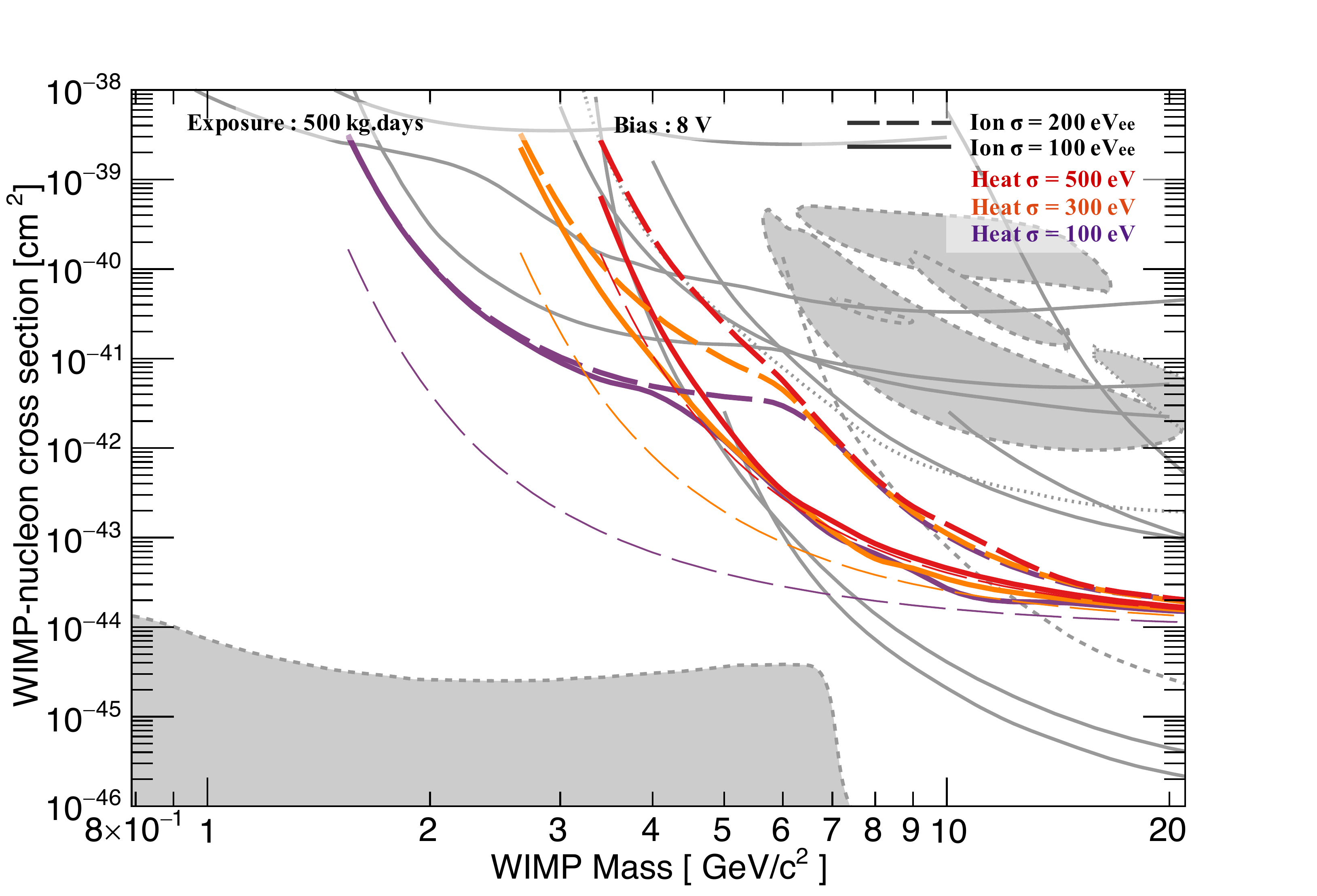} 
\includegraphics[width=0.49\textwidth]{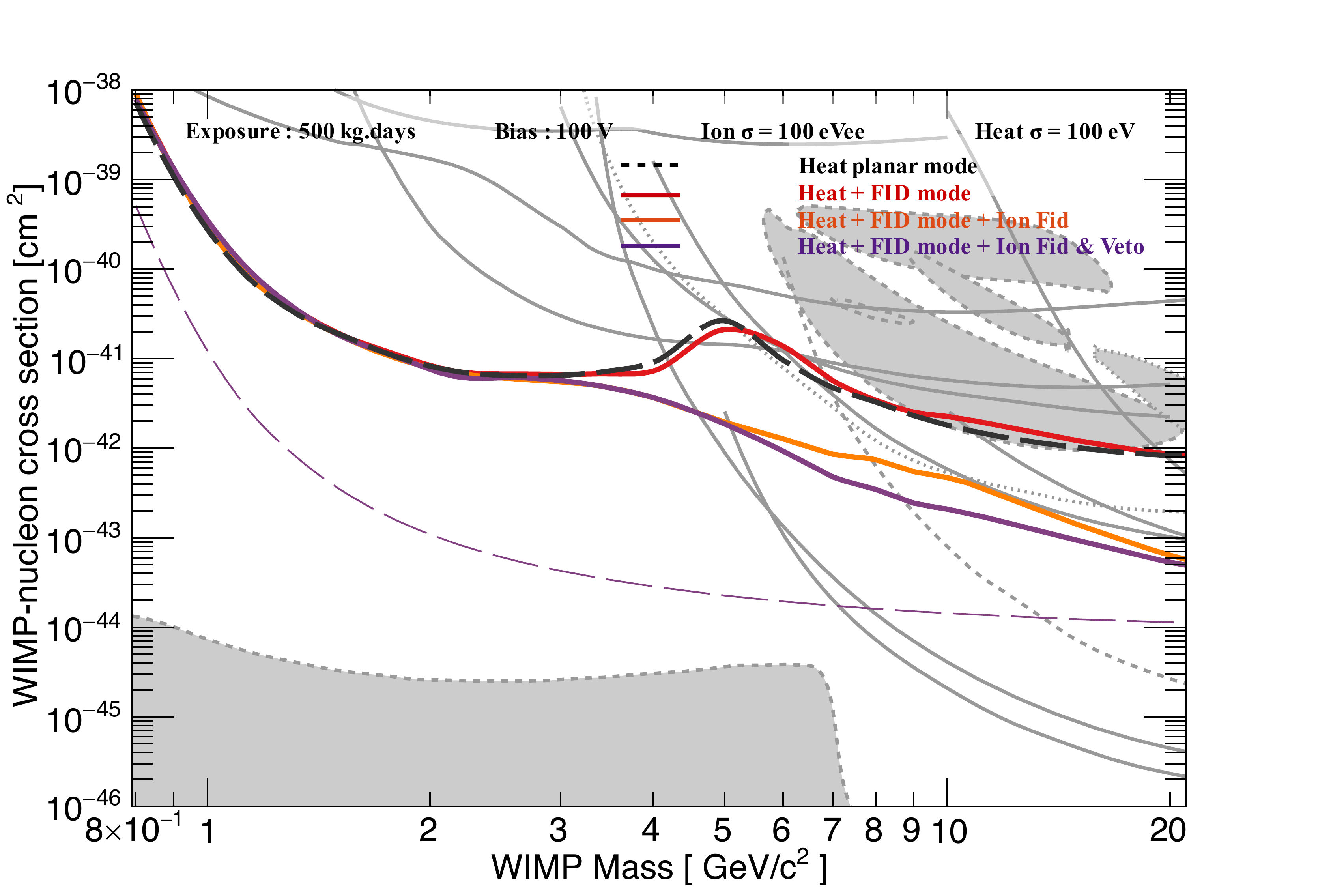} 
\includegraphics[width=0.49\textwidth]{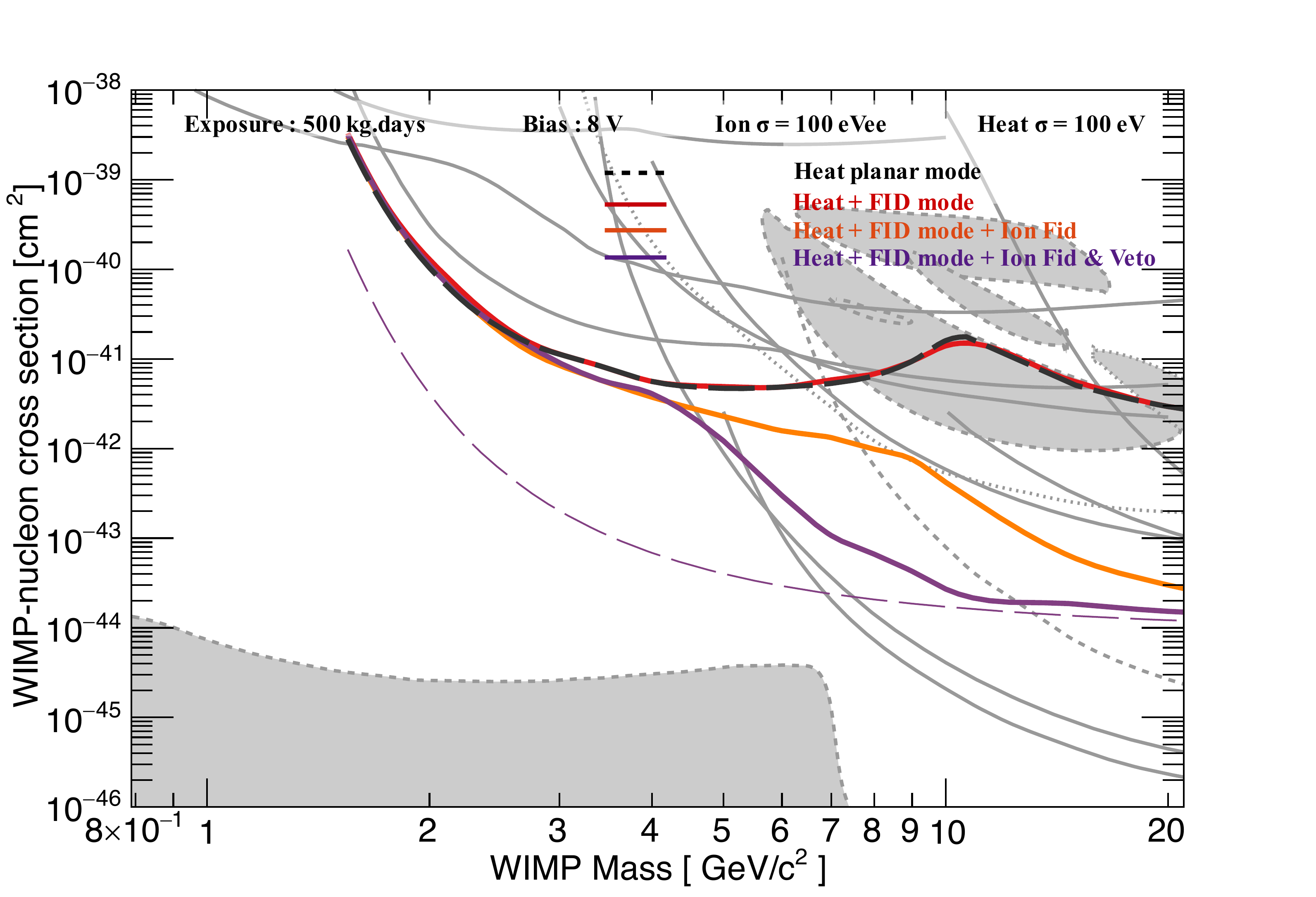}
\caption{\label{fig:allpanels}Effect of various experimental parameters on the sensitivity derived from the likelihood analysis for an exposure of 500 $\mathrm{kg\cdot d}$ at both 100~V (left panels) and 8~V (right panels). All upper limits, contours and regions plotted in gray in this figure are those described in Fig.~\ref{fig:SotA2016}.\\
Top panels: impact of backgrounds for fixed heat and ionization resolutions ($\sigma_{E_{heat}}=100\;\mathrm{eV}$, $\sigma_{E_{fid}}=100\;\mathrm{eV_{ee}}$). The solid red line corresponds to the exclusion limit when all backgrounds are included. The dashed blue line indicates the background-free sensitivity. The other limits are derived by considering each background separately as indicated by the color code. \\
Middle panels: impact of varying heat and ionization energy resolutions considering all background components. Heat energy resolution values are distinguished by their color code: red, orange and purple for $\sigma_{E_{heat}}=500\;\mathrm{eV}$, $\sigma_{E_{heat}}=300\;\mathrm{eV}$ and $\sigma_{E_{heat}}=100\;\mathrm{eV}$, respectively, while dashed and solid lines correspond to $\sigma_{E_{fid}}=200$ and $100\;\mathrm{eV_{ee}}$, respectively. The thin dashed lines correspond to the bakground-free sensitivity, with the same color code as used for $\mathrm{E_{heat}}$.\\
Bottom panels: impact of different detector designs considering all the background components for fixed heat and ionization resolutions ($\sigma_{E_{heat}}=100\;\mathrm{eV}$, $\sigma_{E_{fid}}=100\;\mathrm{eV_{ee}}$). The solid purple lines refer to the standard FID detector design and the orange ones correspond to the same design without the ability to read veto channels. The red lines correspond to an FID detector only reading the heat channel, with still two distinct collection biases ($V_{surf}$ and $V_{fid}$). Finally the dashed black line refers to a detector in coplanar mode only equipped with one heat channel. The purple thin dashed line corresponds to the bakground-free sensitivity associated to the standard FID detector design.}
\end{figure*}
Achievable sensitivities will be affected by these two opposite effects. In Fig.~\ref{fig:MultiVolt}, we show how the projected sensitivies vary by increasing the bias voltage from 10 V to 100 V while keeping all other detector characteristics the same: the whole set of EDELWEISS-III backgrounds with a fixed exposure of 500~$\mathrm{kg\cdot d}$, $\sigma_{E_{heat}}=100$~eV, and $\sigma_{E_{fid}}=100\;\mathrm{eV_{ee}}$. We observe that the WIMP mass range is clearly splitted in two regions, below and above 4~GeV/c$^2$, which are the so-called low-mass and intermediate-mass regions, respectively:\newline
- above 4~GeV/c$^2$, we observe a loss of the sensitivity with increasing bias attributable to a decreasing discrimination power. The latter is particularly marked for WIMP masses around 10~GeV/c$^2$ with a sensitivity reduction by almost one order of magnitude when varying $V_{fid}$ from 10~V to 100~V.\newline 
- below 4 GeV/c$^2$, the signal amplification provides sensitivity to lower WIMP masses since both the trigger and analysis thresholds depend on the heat resolution. Furthermore, for a given WIMP mass, high biases lead to much more stringent limits. A similar improvement of both the background-free sensitivities and the exclusion limits in presence of backgrounds is observed, which could  indicate that below 4 GeV/c$^2$, the discrimination power is not affected anymore by the use of high biases. This effect will be further discussed in section~\ref{sub:readout}.

In order to properly study the effect of various experimental conditions on the sensitivity to both intermediate- and low-mass regions for WIMPs, only two extreme $V_{fid}$ conditions will be considered in the following studies, which are $V_{fid} =100$~V for the high value (left panels of Fig.~\ref{fig:allpanels}) and $V_{fid} =8$~V for the low one (right panels of Fig.~\ref{fig:allpanels}).
      \subsection{Impact of backgrounds}
\label{sub:backgroundimpact}
The impact of the main background contributions on the exclusion sensitivity has been studied at fixed energy resolutions ($\sigma_{E_{heat}}=100\;\mathrm{eV}$, $\sigma_{E_{fid}}=100\;\mathrm{eV_{ee}}$), either taking into account all backgrounds as described in section~\ref{sub:backgroundmodel} or for a background-free experiment. Between these two extreme cases, each background type with substantial event rate (see Table~\ref{tab:rates}) has been considered separately: Compton, tritium, beta, lead and heat-only events. The impact of $^8$B neutrino or neutron background is negligible on the sensitvity, whereas cosmogenic X-ray lines have been easily subtracted thanks to the good energy resolutions used for these projections. The upper limit associated to each considered background is derived by setting to zero all the other background components. \\
As illustrated on the top panels of Fig.~\ref{fig:allpanels}, which show the exclusion limits associated to the different background types 
for an exposure of 500 $\mathrm{kg\cdot d}$, 
the heat-only background clearly dominates for both voltage bias conditions 
below 5 GeV/c$^2$. \\
Considering WIMP searches at 100~V, suppression of the heat-only background would increase the sensitivity by more than one order of magnitude in the low-mass region. Also for this $V_{fid} = 100$~V value, Compton and tritium backgrounds are the dominating ones for the intermediate-mass region, above $\sim 10$~GeV/c$^2$. This is due to the Neganov-Luke effect dominating the heat signal which implies that the discrimination between electronic and nuclear recoils gets dramatically reduced.
       \subsection{Effect of energy resolutions}
\label{sub:resolutions}
The impact of varying both heat and ionization energy resolutions has been studied either taking into account all backgrounds or for a background-free experiment, again for an exposure of 500 $\mathrm{kg\cdot d}$. Three cases have been considered for the heat energy resolutions  ($\sigma_{E_{heat}}=500\;\mathrm{eV}$, $300\;\mathrm{eV}$ and $100\;\mathrm{eV}$) and two values used for the ionization energy resolutions ($\sigma_{E_{fid}}=200\;\mathrm{eV_{ee}}$ or $100\;\mathrm{eV_{ee}}$), varying from the current resolution values of the EDELWEISS-III experiment to the expected achievable ones. Results are presented in the middle panels of Fig.~\ref{fig:allpanels}.\\
Let us first consider the limits obtained when operating at $V_{fid} =100$~V (left panel). Reducing the heat energy resolution $\sigma_{E_{heat}}$ leads to reduced thresholds and therefore to an improved sensitivity to low-mass WIMPs, while improving the ionization resolution $\sigma_{E_{fid}}$ has no effect on this sensitivity as shown by the overlap of the solid and dashed lines below 4 GeV/c$^2$. Above this WIMP mass value, the behaviour is the exact opposite: reducing the heat resolution doesn't increase the sensitivity in contrast with improving ionization resolution. Lowering $\sigma_{E_{fid}}$ allows attenuating the loss of discrimination power originating from high voltage biases by half an order of magnitude.\\
Considering now the limits derived from the $V_{fid} = 8$~V scenario the conclusions are nearly the same than at 100~V: reducing the ionization resolution $\sigma_{E_{fid}}$ from 200~$\mathrm{eV_{ee}}$ to 100~$\mathrm{eV_{ee}}$ is extremely favourable in the intermediate-mass region. This is essentially due to both the resulting reduced overlap of the heat-only events within the ROI and to the increased surface event rejection power. This upgrade in ionization has, however, no impact on the sensitivity to lower WIMP masses. Taking into account the possible improvement of the heat energy resolution down to $\sigma_{E_{heat}}=100\;\mathrm{eV}$, it would also provide sensitivity to lower WIMP masses. However the discrimination power is not improved in most of the WIMP mass range and especially in the intermediate-mass region. This lack of improvement is clearly shown from the limits unchanged for WIMP masses above 4~GeV/c$^2$ at $\sigma_{E_{fid}}=100\;\mathrm{eV_{ee}}$ and above 6~GeV/c$^2$ at $\sigma_{E_{fid}}=200\;\mathrm{eV_{ee}}$, respectively.\\
To conclude on the energy resolution effect, reducing the heat energy resolution $\sigma_{E_{heat}}$ down to 100 eV leads to reduced thresholds and therefore to an improved sensitivity to low-mass WIMPs, especially at $V_{fid} =100$~V.

      \subsection{Detector design and readout channels}
\label{sub:readout}
One can wonder what is the contribution of the double readout to the sensitivity in the low-mass region. Indeed exclusion limits derived at $\sigma_{E_{fid}}=100\;\mathrm{eV_{ee}}$ and $\sigma_{E_{fid}}=200\;\mathrm{eV_{ee}}$ present an extreme similarity for both $V_{fid}$ values at low WIMP mass as shown with the previous study (see the middle panels of Fig.~\ref{fig:allpanels}). 
To go further, a performance comparison has been carried out using four different detector designs which have or not the capability to produce the double readout.\\
Resulting exclusion limits are presented on the bottom panels of Fig.~\ref{fig:allpanels} at 100~V (left) and 8~V (right). The solid purple lines refer to the standard FID detectors, which have been considered to compute all previous limits shown from Fig.~\ref{fig:LikeBDT} to Fig.~\ref{fig:allpanels} (middle panels). The solid orange lines correspond to the same design without the ability to read veto channels and therefore to reject surface events. The red lines indicate the expected performance from an FID detector only reading the heat channel, with still two distinct collection biases for surface events ($V_{surf}$) and fiducial events ($V_{fid}$). Finally, in dashed black, we report the expected results from the simplest detector design: a detector in coplanar mode only equipped with one heat channel. 
All the simulations are computed considering an exposure of 500~$\mathrm{kg\cdot d}$, either all the current background components or the background-free sensitivity, and expected energy resolutions of $\sigma_{E_{fid}}= 100\;\mathrm{eV_{ee}}$ in ionization and $\sigma_{E_{heat}} = 100\;\mathrm{eV}$ in heat. \\
We clearly identify from the overlap of the four limits at 8~V and 100~V below 4 GeV/c$^2$ mass WIMP that the most simple detector design is enough to probe such low-mass region. As discussed in section~\ref{sub:backgroundimpact}, the main limitation here is the heat-only background for which only an improvement of the ionization energy resolution or a reduction of the associated event rate could increase the sensitivity for a given exposure. Even in case this background would be significantly reduced or suppressed, conclusions would be unchanged as the beta background would take over and neither the surface rejection capability, nor the ionization/heat based discrimination is still effective at such low mass. However, once the discrimination starts to be feasible with respect to the assumed energy resolutions, 
there is a significant gain confered by the double readout, especially at 8~V that is furthermore the bias condition to consider to probe the intermediate-mass region. Also, the limits show that measuring ionization allows for spectral shape discrimination, which is particularly important to differentiate 5~GeV/c$^2$ (10 GeV/c$^2$) WIMPs at 100~V (8~V) from heat-only events as the similarity of their heat energy spectra is responsible for the visible bumps in sensitivity for the two simplest detector designs considered. \\
In terms of detector design, the discrimination power of the double readout is needed only to obtain a good sensitivity in the intermediate-mass WIMP region.
       \section{Prospects}
\label{sec:prospects}
       \subsection{EDELWEISS-III low mass projections}
\label{subsec:prospects-EDW2018}
\begin{figure*}[h]
\begin{center}
\includegraphics[width=0.80\textwidth]{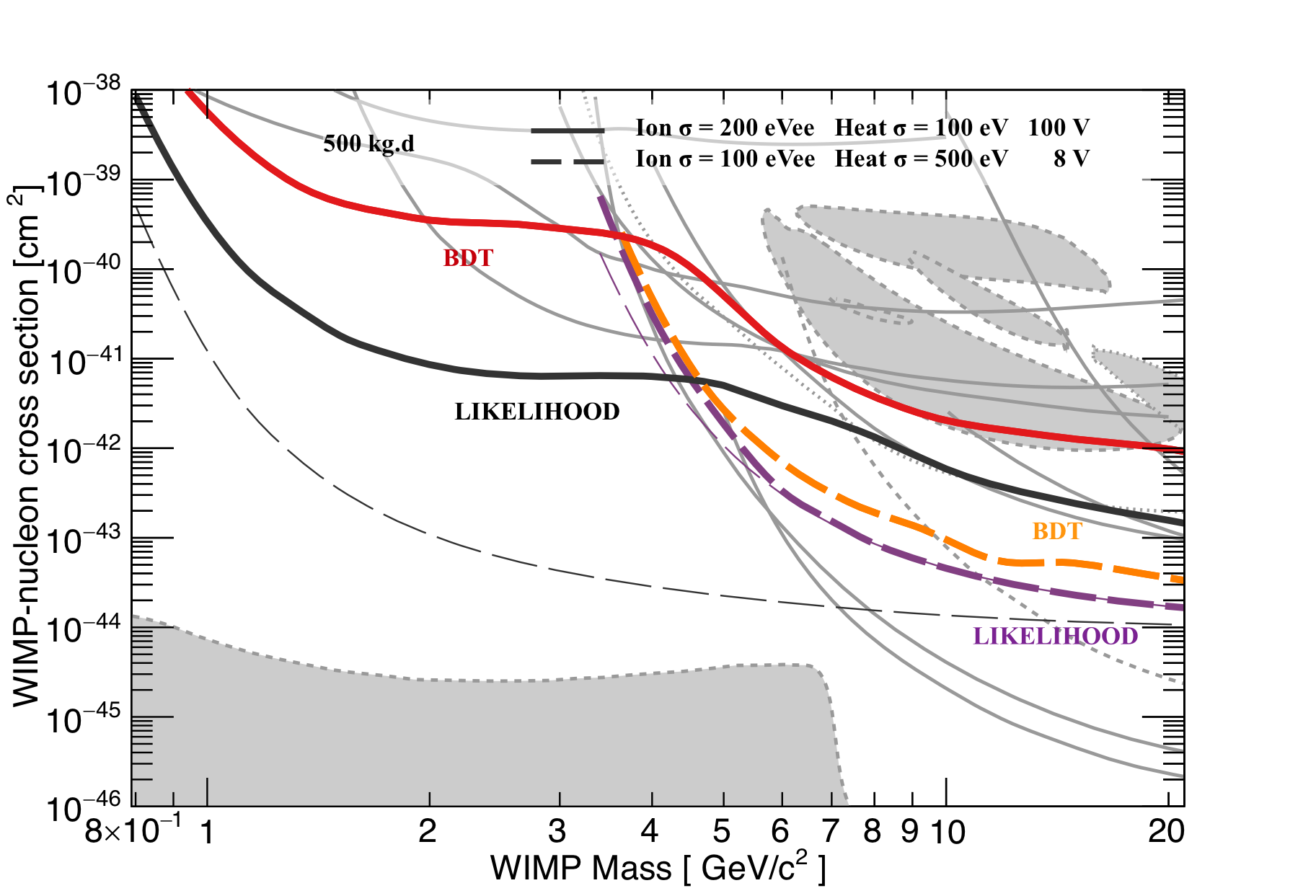}
\end{center}
\caption{\label{fig:Projection}EDELWEISS-III projected sensitivities considering that expected R\&D upgrades will be achieved either for ionization or heat resolutions, with current LSM setup and background budget. Exclusion limits are derived from both boosted decision tree (BDT) and profile likelihood ratio approaches and for the two extreme bias voltage conditions (8~V and 100~V). For both $V_{fid}$ values a likelihood analysis gives better limits than a BDT one. 
Below WIMP masses of 4-5 GeV/c$^2$, the best sensitivity is obtained by lowering the thresholds, with the Luke-Neganov boost corresponding to a  bias voltage of 100~V, keeping current ionization resolution $\sigma_{E_{fid}}=\mathrm{200\;eV_{ee}}$ and improving heat resolution to $\sigma_{E_{heat}}=\mathrm{100\;eV}$. Official EDELWEISS-III low-mass projected sensitivity is thus given by the black solid line exclusion limit.  The background-free sensitivity is shown in thin dashed lines. 
All upper limits, contours and regions plotted in gray in this figure are those described in Fig.~\ref{fig:SotA2016}.\\}
\end{figure*}

Fig.~\ref{fig:Projection} presents the two major scenarios for near future low-mass WIMP search with EDELWEISS-III, considering efforts have been put on the R\&D, aiming at improving at least one of the energy resolutions, either for heat or for ionization signals. Sensitivities have been computed with both BDT and likelihood methods for a total exposure of 500~$\mathrm{kg\cdot d}$ with our current background levels and setup at the LSM. \\
Improving ionization resolution could be done through the implementation of High Electron Mobility Transistors (HEMT) to replace Junction Field Effect Transistors (JFET) used for charge measurements on the Al electrodes collecting electron-hole pairs~\cite{performance-paper}. As shown in~\cite{HEMT}, a calibrated baseline energy resolution of 91~eV$_{ee}$ has been already achieved with a HEMT-based charge amplifier
coupled to a live CDMS-II detector. Thus, the next R\&D step could be the coupling of this charge amplifier to an EDELWEISS detector with the goal of obtaining the ionization resolution $\sigma_{E_{fid}}=\mathrm{100\;eV_{ee}}$. 
Concerning heat resolution improvement, dedicated R\&D is also in progress on baseline performance with the achievable objective of reaching $\sigma_{E_{heat}}=\mathrm{100\;eV}$:  a coherent thermal model has been constructed as described in~\cite{ThermalModel} and is used to extract relevant parameters of the heat signal in order to build new thermal sensors which would provide the expected heat energy resolution improvement. It could lead to nuclear recoil energy thresholds ranging from 400 to 100 eV$_{nr}$, depending on the applied bias voltage ($V_{fid}$) across the crystal.\\
Considering that expected values for either heat or ionization baseline resolutions have been achieved, the two studied scenarios are taking either usual low bias voltages or high ones. Exclusion curves corresponding to the first scenario are computed using  $\sigma_{E_{fid}}=\mathrm{100\;eV_{ee}}$ for ionization instead of $\sigma_{E_{fid}}=\mathrm{200\;eV_{ee}}$,  keeping current heat resolution performance ($\sigma_{E_{heat}}=\mathrm{500\;eV}$) and applying low bias voltage at $V_{fid} = 8$~V, which allows some discrimination performance. The second scenario uses the possibility to deal with current ionization resolution at $\mathrm{200\;eV_{ee}}$ and prioritizes the R\&D aiming at lowering the thresholds. The corresponding exclusion curves have been computed assuming an achieved performance of $\sigma_{E_{heat}}=\mathrm{100\;eV}$ and working at high $V_{fid}= 100$~V. \\
Both analysis methods provide similar results in the first scenario at 8~V (orange and purple dashed lines in Fig.~\ref{fig:Projection}). Keeping current heat resolution $\sigma_{E_{heat}}=\mathrm{500\;eV}$ and low bias voltage of 8~V for discrimination performance, and improving only ionization resolution to $\sigma_{E_{fid}}=\mathrm{100\;eV_{ee}}$ is not a good solution since obtained exclusion curves were not competitive with current results from noble liquid experiments.
For $V_{fid} = 100$~V a likelihood analysis gives better limits than a BDT one (black and red solid lines in Fig.~\ref{fig:Projection}).
Below WIMP mass of 4-5 GeV/c$^2$, the best sensitivity is obtained by lowering the thresholds, with the Luke-Neganov boost corresponding to a  bias voltage of 100~V, keeping actual ionization resolution $\sigma_{E_{fid}}=\mathrm{200\;eV_{ee}}$ and improving heat resolution to $\sigma_{E_{heat}}=\mathrm{100\;eV}$. 

Hence, prioritizing HEMT's implementation to achieve expected low ionization resolution would not be an issue even if the background model was not precise enough to perform a likelihood analysis. However, no improvement is expected with respect to current sensitivities already achieved by other experiments (see gray lines of Fig.~\ref{fig:Projection}).
In the second scenario at 100~V, where lower thresholds are achieved both by improving heat resolution and boosting the Neganov-Luke effect (red and black solid lines on Fig.~\ref{fig:Projection}), it is possible to put new constraints on almost half of the remaining uncovered parameter space region, though a likelihood analysis is fundamental to fulfill this purpose. \\
Below WIMP masses of $\sim 5$~GeV/c$^2$ the best sensitivity is obtained with the Luke-Neganov boost corresponding to a  bias voltage of 100~V to lower the energy threshold. This is the reason why the EDELWEISS collaboration is now focusing on R\&D to put high voltage biases on detectors, in addition to improving heat and ionization baseline resolutions. Thus the official EDELWEISS-III low mass projection is presented on Fig.~\ref{fig:Projection} in black solid line. 


         \subsection{EDELWEISS 100 kg-scale}
\label{subsec:prospects-EDW100kg}
\begin{figure*}[h]
\begin{center}
\includegraphics[width=0.80\textwidth]{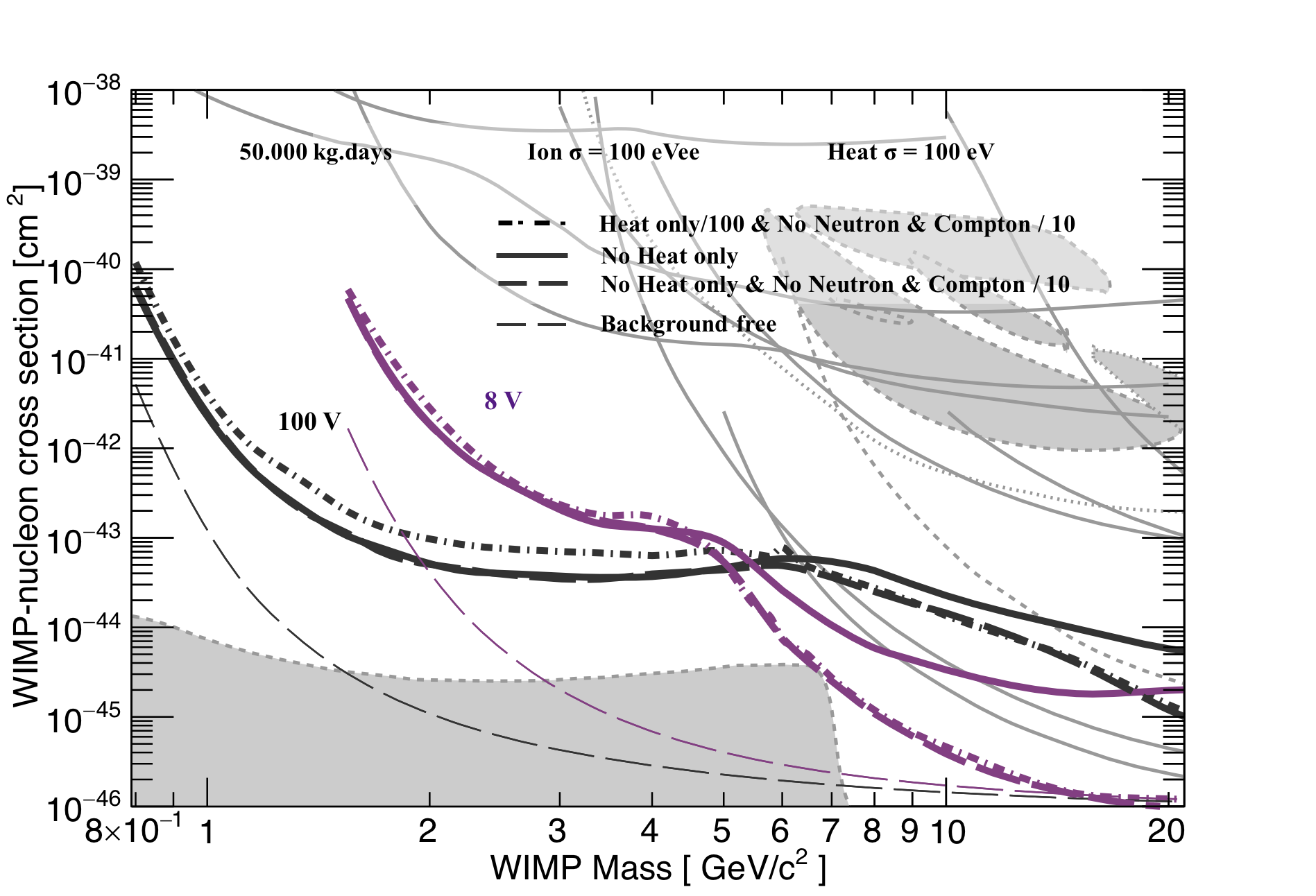}
\end{center}
\caption{\label{fig:Projection2}Projected sensitivities for a large exposure of  $50\,000$ $\mathrm{kg\cdot d}$ in the context of a hundred-kg-scale EDELWEISS-like experiment with strongly improved background levels and R\&D upgrade performance achieved, with baseline resolutions of $\sigma_{E_{heat}}=\mathrm{100\;eV}$ in heat and $\sigma_{E_{fid}}=\mathrm{100\;eV_{ee}}$ in ionization.
Limits are computed using a likelihood analysis at 8~V (purple) and 100~V (black) assuming a suppression of the heat-only background (solid line), and no more neutron background associated with a reduction of the Compton background by a factor 10 (thick dashed line). The background-free sensitivity is shown in thin dashed lines. All upper limits, contours and regions plotted in gray in this figure are those described in Fig.~\ref{fig:SotA2016}.}
\end{figure*}

Looking further ahead, one considers the requirements to approach the neutrino floor~\cite{Billard}, which corresponds to the coherent scattering of neutrinos from several astrophysical sources as solar $^8$B neutrinos. This could produce background by almost perfectly mimicking a WIMP signal (see section~\ref{sub:backgroundmodel}).
Fig.~\ref{fig:Projection2} shows sensitivity projections derived from the likelihood analysis for a large exposure of $50\,000$ $\mathrm{kg\cdot d}$ and resolutions of $\sigma_{E_{heat}}=\mathrm{100\;eV}$ in heat and $\sigma_{E_{fid}}=\mathrm{100\;eV_{ee}}$ in ionization. Limits are computed for both 8~V and 100~V bias voltages and plotted in purple and black, respectively.  
Solid lines of Fig.~\ref{fig:Projection2} correspond to the expected limits achievable considering the current EDELWEISS background budget, with the exception of  heat-only events, which are supposed to be completely suppressed. Thick dashed lines (dot-dashed lines) are obtained assuming not only no more heat-only events (a reduction of heat-only events by a factor 100), but also no more neutrons and a reduction of the Compton background by a factor 10. These latters could be obtained by putting upgraded EDELWEISS detectors in a dedicated environment with high radiopurity level, such as SNOLAB\footnote{see https://www.snolab.ca/} for example~\cite{SuperCDMS-projections, EURECA}. Remaining backgrounds are then surface events and events from coherent neutrino-nucleus scattering induced by solar $^8$B neutrinos. The background-free sensitivity is shown in thin dashed lines. \\
As illustrated by the dashed and dot-dashed lines, in purple and black for 8~V and 100~V, respectively, reducing the Compton and neutron backgrounds will improve the sensitivity only for WIMP masses above 5-6 GeV/c$^2$, which is the mass region where large-scale LXe 
dual-phase TPC detectors such as Xenon1T~\cite{XENON1T} and LZ~\cite{LZ}
may dominate over cryogenic experiments. \\
We note two features of the $V_{fid} = 100$~V scenario, associated to the black lines of Fig.~\ref{fig:Projection2}:\\
- For WIMP masses above 6 GeV/c$^2$, suppression of heat-only events would not be the major issue: the best upper limits are obtained by reducing neutron and Compton backgrounds. The limits obtained simulating either a reduction of heat-only events (black dot-dashed line) or a total suppression of this background (black dashed line) are overlapping on Fig.~\ref{fig:Projection2}. \\
- On the contrary, below 6 GeV/c$^2$ the heat-only background dominates all other backgrounds and the best sensitivities (black solid and dashed lines) require that it be strongly suppressed. The remaining difference between these two curves and the background-free sensitivity (thin black dashed line) is due to surface backgrounds. The projections conservatively assume no improvement relative to currently observed levels. Approaching the neutrino floor at $V_{fid} = 100$~V would require an order of magnitude improvement on both the selection of material in direct contact with the germanium crystal and the cleaning of the detector and support surfaces, reaching the levels quoted in~\cite{SuperCDMS-projections}. \\
In the second scenario, with $V_{fid} = 8$~V (purple lines of Fig.~\ref{fig:Projection2}), the best upper limits are obtained for the whole WIMP mass range by reducing neutron and Compton backgrounds, as shown by the identical curves obtained, either with a reduction of heat-only events (purple dot-dashed line) or a total suppression of this background (purple dashed line). It is thus clear that a more radiopure environment will be needed to take advantage of the potential of the EDELWEISS detectors. \\
The best sensitivity will be achieved above 5~GeV/c$^2$ with $V_{fid} = 8$~V bias voltage put on FID upgraded detectors in a SNOLAB-like environment, thanks to their discrimination power. However none of the scenarios will allow to reach the neutrino floor. \\
It is worth mentioning at this point that these projected sensitivities for low-mass WIMP search are as good (or even better) than the already published ones by other cryogenic experiments such as SuperCDMS@SNOLAB~\cite{SuperCDMS-projections} and CRESST-III~\cite{CRESST-projections}.

         \subsection{Entering the $^8$B region}
\label{subsec:prospects-EDWbore8}
\begin{figure*}[htb]
\includegraphics[width=0.80\textwidth]{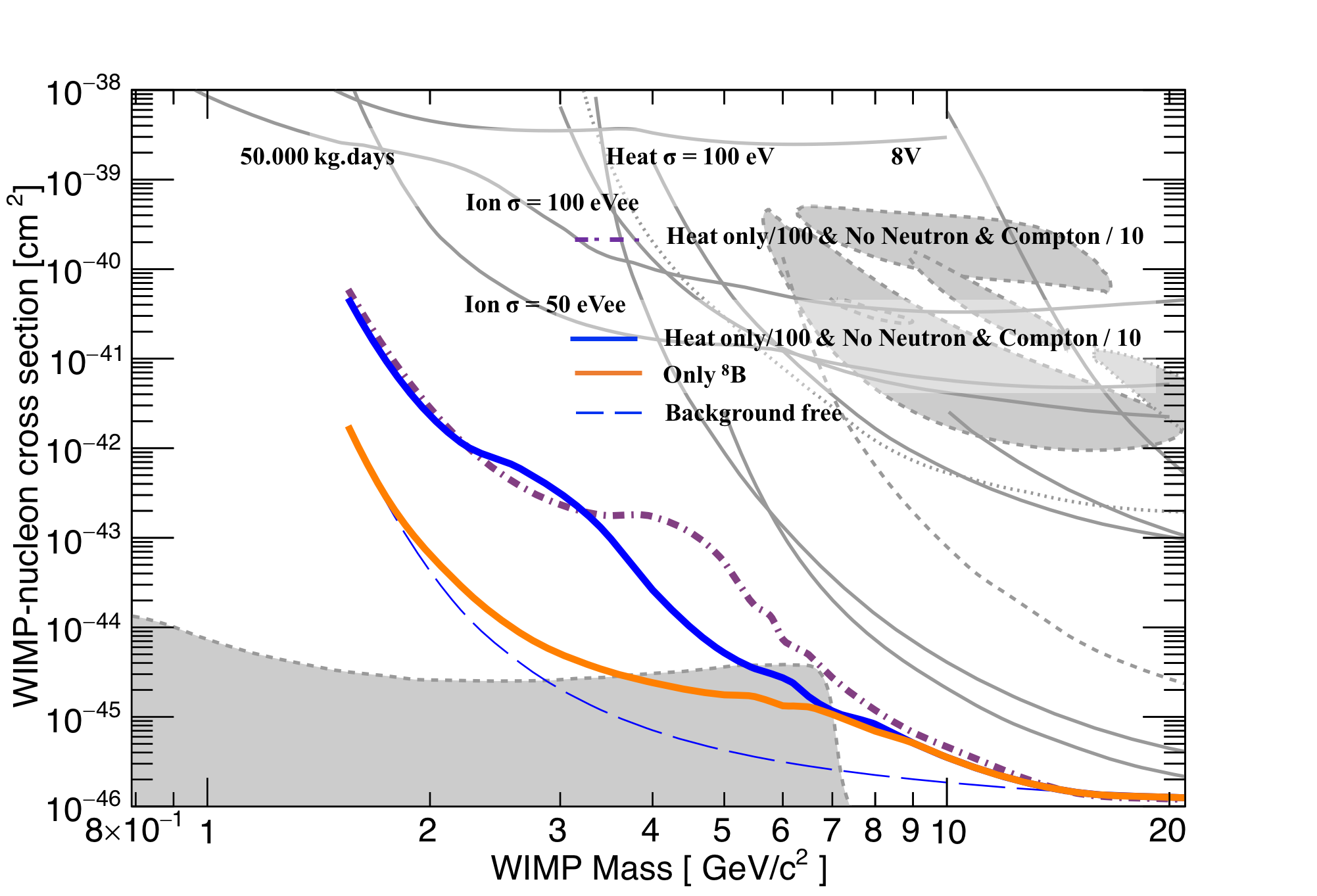}
\caption{\label{fig:limit8B}Projected sensitivities computed using a likelihood analysis at 8~V for a large exposure of  $50\,000$ $\mathrm{kg\cdot d}$ in the context of a hundred-kg-scale EDELWEISS-like experiment, considering R\&D upgrade performed on heat baseline resolution, with $\sigma_{E_{heat}}=\mathrm{100\;eV}$, and assuming different conditions of background reduction. The purple dot-dashed line upper limit, which is obtained with $\sigma_{E_{fid}}=\mathrm{100\;eV_{ee}}$, considering a reduction of heat-only background by a factor 100, no more neutron background and a reduction of the Compton background by a factor 10, obtained by putting upgrading EDELWEISS detectors in a high-purity level dedicated environment, has been already shown in Fig.~\ref{fig:Projection2}. Its purpose is to guide the eye. \\
The projected sensitivities in solid and thin dashed lines shown on this figure are obtained assuming a more drastic improvement on ionization resolution, with $\sigma_{E_{fid}}=\mathrm{50\;eV_{ee}}$. Solid line upper limits correspond to different background reduction: the blue one is obtained with the same background conditions than for purple dot-dashed line whereas the orange one is performed removing all backgrounds including surface events, except the one due to $^8$B neutrinos. Blue thin dashed line represents the background-free sensitivity. 
All upper limits, contours and regions plotted in gray in this figure are those described in Fig.~\ref{fig:SotA2016}.}
\end{figure*}

\begin{figure*}[htb]
\includegraphics[width=0.8\textwidth]{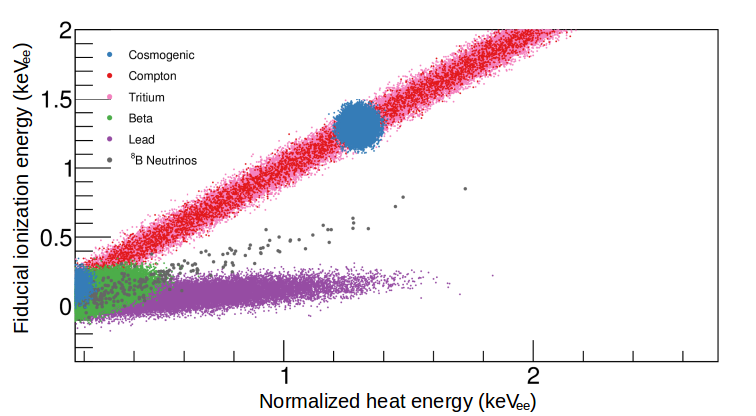}
\caption{\label{fig:ion-vs-heat}Simulation of $^8$B coherent neutrino-nucleus scattering signal and background events, assuming current EDELWEISS-III background budget, with the exception of no more heat-only, no neutron events, and a reduction of the Compton background by a factor 10. These simulations are computed for an exposure of 1\,000 $\mathrm{kg\cdot yr}$, at 8~V, with $\sigma_{E_{heat}}=\mathrm{100\;eV}$ and $\sigma_{E_{fid}}=\mathrm{50\;eV_{ee}}$ baseline energy resolutions. The separation of $^8$B signal (medium gray dots) from background events is clearly shown in the plane of fiducial ionization energy versus normalized heat energy, in keV$_{ee}$.}
\end{figure*}

\begin{figure*}[htb]
\includegraphics[width=0.75\textwidth]{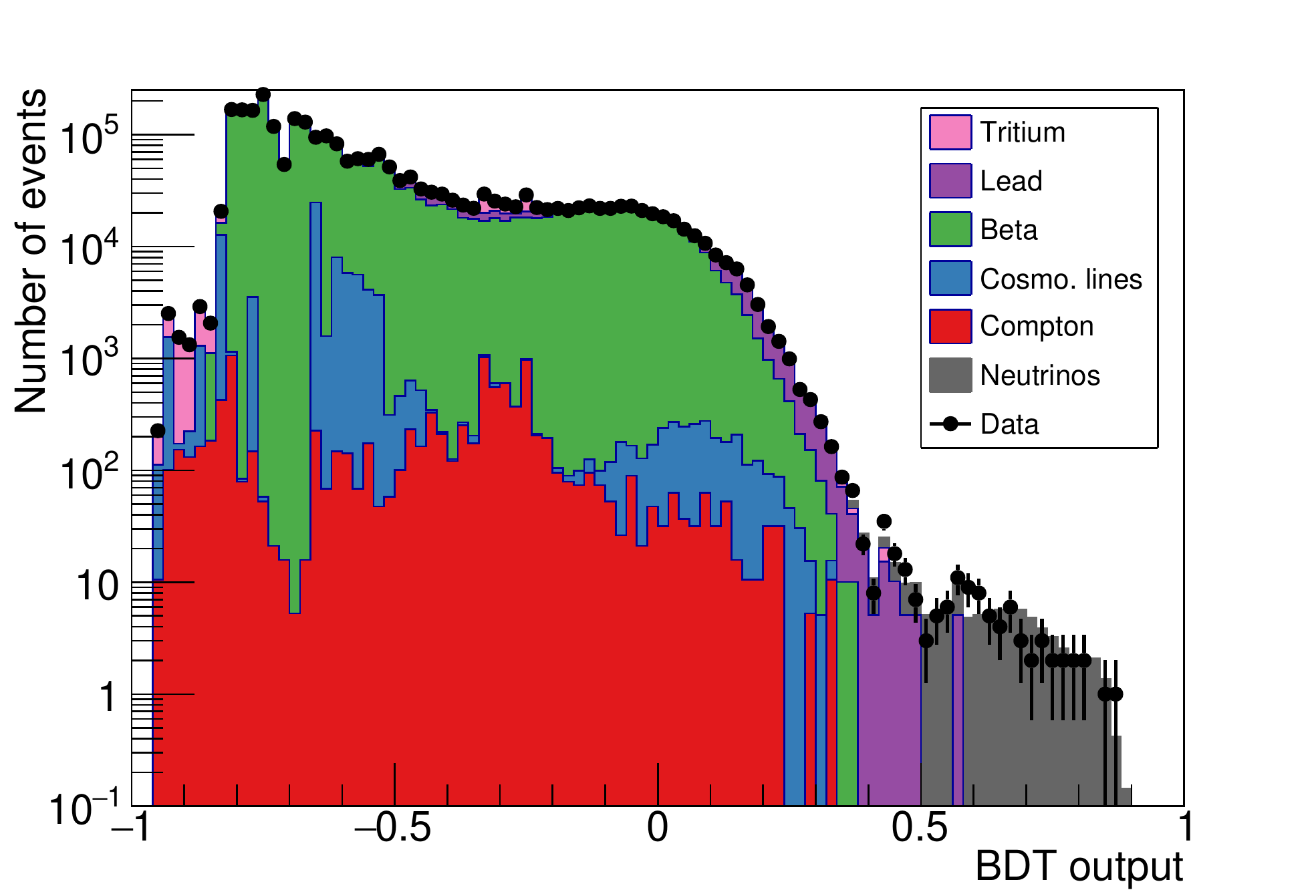}
\caption{\label{fig:BDT-score}BDT distribution simulated for a $^8$B neutrino signal, obtained for an exposure of 1\,000 $\mathrm{kg\cdot yr}$, at 8~V, with heat baseline resolution of $\sigma_{E_{heat}}=\mathrm{100\;eV}$ and ionization baseline resolution of $\sigma_{E_{fid}}=\mathrm{50\;eV_{ee}}$. Colored distributions are associated to the different background models. The expected $^8$B neutrino signal is represented in gray. The BDT distribution of the simulated data in the ROI corresponds to the black dots. Above a BDT score cut of 0.5, a clear signal of 78 $^8$B neutrino events is obtained after discrimination.}
\end{figure*}

The next step to improve exclusion limits on the spin-independent WIMP-nucleon cross-section is to reach and eventually enter the neutrino floor. As discussed in the previous section, the easier way would be the use of $V_{fid} = 8$~V bias voltage to benefit from the powerful discrimination between electronic and nuclear recoils of the FID detectors. 
We consider in Fig.~\ref{fig:limit8B} the projected sensitivity curves for a large exposure of  $50\,000$ $\mathrm{kg\cdot d}$ in a dedicated environment with high radiopurity level, considering that the R\&D upgrade has been achieved on heat baseline resolution, with $\sigma_{E_{heat}}=\mathrm{100\;eV}$. Entering the neutrino floor could be realized even keeping a reduced number of heat-only events, by improving more drastically the ionization baseline resolution from $\sigma_{E_{fid}}=\mathrm{100\;eV_{ee}}$ (purple dot-dashed line) to $\sigma_{E_{fid}}=\mathrm{50\;eV_{ee}}$ (blue solid line). \\
However, looking at the orange solid line of Fig.~\ref{fig:limit8B} corresponding to the sensitivity when removing all remaining backgrounds, with the exception of the events due to coherent $^8$B neutrino-nucleus scattering, the associated upper limit stays far from the background-free sensitivity. Nevertheless, such a very good ionization energy resolution could be used to study the $^8$B coherent neutrino-nucleus scattering as a signal instead of a background. Upgraded FID bolometers have the good design for this study since solar $^8$B neutrino spectral shape (as shown in Fig.~\ref{fig:recoilenergybackgrounds}) is really similar to the one of a 6 GeV/c$^2$~WIMP signal. As a confirmation, Fig.~\ref{fig:ion-vs-heat} shows fiducial ionization energy as a function of normalized heat energy for various types of background events and for the $^8$B coherent neutrino-nucleus scattering considered as the searched signal (black dots). This plot has been obtained by simulating an exposure of 1\,000 $\mathrm{kg\cdot yr}$, under the same background hypothesis than presented in section~\ref{subsec:prospects-EDW100kg} (see dashed lines of Fig.~\ref{fig:Projection2}): no more heat-only events, no more neutrons, a reduction of the Compton background by a factor 10, keeping all other EDELWEISS-III current background levels.  Also baseline energy resolutions of $\sigma_{E_{heat}}=\mathrm{100\;eV}$ and $\sigma_{E_{fid}}=\mathrm{50\;eV_{ee}}$ have been used to perform the plot, with the FID detector design and using fiducial cuts as described in section~\ref{sub:detmodel}. In order to  demonstrate the discrimination power, the $^8$B neutrino signal is represented with medium gray dots on Fig.~\ref{fig:ion-vs-heat} 
The separation of this $^8$B signal from background events is obtained thanks to the very good ionization resolution of $\sigma_{E_{fid}}=\mathrm{50\;eV_{ee}}$ allowing a powerful spectral separation as clearly shown in the figure. For reference, the number $N(^8{\rm B})$ of expected $^8$B neutrino events above both $E_{fid} = 0.2$~keV$_{ee}$ and $\tilde{E}_{heat} = 0.5$~keV$_{ee}$, where there is a clear separation with backgrounds, is $N(^8{\rm B}) = 78$  for this exposure of $1\,000$ $\mathrm{kg\cdot yr}$. \\
With these very good energy resolutions, better than or around 10\% at 1~keV$_{ee}$, one can perform a simulated BDT analysis.  As described in section~\ref{sec:bdt}, results of a BDT analysis are given by a unique ouptut BDT score (Eq.~\ref{BDT_output}) which varies from -1 (background-like) to +1 (signal-like). To compute the EDELWEISS-III expected sensitivity to $^8$B neutrinos, the BDT has been trained for a 6~GeV/c$^2$ WIMP mass (with a spectral shape similar to the one of a $^8$B neutrino) and experimental condition using $10^6$ events generated by Monte Carlo in the 3D-space ($E_{fid},E_{veto},\tilde{E}_{heat}$) according to signal and background models. The BDT analysis is performed with the same baseline energy resolutions 
and background model 
than for Fig.~\ref{fig:ion-vs-heat}, and no fiducial cut is used.
As shown in Fig.~\ref{fig:BDT-score}, which gives the number of events as a function of the BDT output score for an exposure of 1\,000~$\mathrm{kg\cdot yr}$, using a BDT score cut of 0.50 would give a clear signal of 78 $^8$B events after discrimination\footnote{In 100~$\mathrm{kg\cdot yr}$, there is still a signal of eight $^8$B events after discrimination.}. It could pave the way for a detailed measurement of this important, and yet to be observed, signal from $^8$B coherent neutrino-nucleus scattering which will provide a probe for new physics in the low-energy neutrino sector~\cite{Bertuzzo:2017tuf, Cerdeno:2016sfi, Billard:2014yka}.

         \section{Conclusion}
The presented study provides the roadmap for the optimization of EDELWEISS detectors for low-mass WIMP searches. Clear differences have been shown in how to optimize the sensitivity to low-mass WIMPs depending on the focused parameter space region. Below $\sim5$~GeV/c$^2$ WIMP masses, only background levels and the threshold value matter such that simple detector designs with good enough heat energy resolutions of typically 100~eV and high voltage biases of 100~V are better suited. In this low-mass region, the likelihood method is compulsory to benefit from background subtraction and spectral shape discrimination, both essential to avoid a quick saturation effect at low exposure. At higher WIMP masses, the double measurement of heat and ionization signals is a huge asset as long as low voltage biases are applied. For the intermediate-mass region, between 5 and 20 GeV/c$^2$, the ionization resolution is a key experimental performance due to relying on both discrimination power and surface event rejection. Finally, for the high-mass region above 20 GeV/c$^2$, only the total exposure and the neutron background level really matter as long as discrimination between electron and nuclear recoils is provided.\\
Different projected sensitivities for the EDELWEISS-III experiment have been presented: the successful ongoing R\&D programme could lead to a world leading sensitivity down to 1~GeV/c$^2$ with an exposure of only 500 $\mathrm{kg\cdot d}$. Reaching the neutrino floor would however require both important heat-only event reduction and successful detector R\&D, combined with a hundred-kilogram scale Ge experiment, which could be part of a wider collaboration effort. Another considered study concerns the possibility to enter this neutrino floor in order to measure the $^8$B neutrino signal: it could be obtained with HEMT technology, if succeeding to reach an ionization baseline resolution of $\sigma_{E_{fid}}=\mathrm{50\;eV_{ee}}$.
\section*{Acknowledgments}
The help of the technical staff of the Laboratoire Souterrain de Modane and the participant laboratories is gratefully acknowledged. The EDELWEISS project is supported in part by the German ministry of science and education (BMBF Verbundforschung ATP Proj.-Nr.05A14VKA), by the Helmholtz Alliance for Astroparticle Physics (HAP), by the French Agence Nationale pour la Recherche and the LabEx Lyon Institute of Origins (ANR-10-LABX-0066) of the Universit\'e de Lyon in the framework Investissements d'Avenir (ANR-11-IDEX-00007), by the LabEx P2IO (ANR-10-LABX-0038) in the framework Investissements d'Avenir (ANR-11-IDEX-0003-01) both managed by the French National Research Agency (ANR), by Science and Technology Facilities Council (UK) and the Russian Foundation for Basic Research (grant No. 07-02-00355-a).
\appendix*
\section*{Appendix}
\label{sec:appendix}
Below we describe the analytic functions used to compute recoil energy spectra associated to the EDELWEISS-III background model described in section~\ref{sub:backgroundmodel}.\\
\noindent $\bullet$ Compton induced electronic recoils are described by a flat spectrum with amplitude $p_0 = 0.1~\mathrm{dru}$, where the differential rate unit is defined as $1~\mathrm{dru} = 1~(\mathrm{event/kg.d.keV})$~\cite{lewin}. \\
\noindent  $\bullet$ Recoil energy spectrum for tritium $\beta$ decays is given by Eq.~\ref{Eq-tritium}, with $p_{j=0,2}$ as parameters: 
\begin{equation}
\label{Eq-tritium}
\dfrac{dR}{dE_r}=\left[p_0\, \left(p_1 - E_r\right)^2 \left(p_2 +E_r \right) \sqrt{E_r^2+2p_2E_r} \right]
\end{equation}
$p_0 = 1.406 \times 10^{-8}\, \mathrm{dru\cdot keV}^{-4}$, $p_1 = 18.6$~keV, $p_2 = 511$~keV.\\
\noindent  $\bullet$  Recoil energy spectrum for surface beta events from the $^{210}$Pb decay chain has been derived by fitting the averaged spectrum in data between 5 and 50~keV with the following function: 
\begin{equation}
\label{Eq-surface-210Pb}
\dfrac{dR}{dE_r}=\left[p_0\, \mathrm{exp}\left(p_1E_r\right)+p_2\,\mathrm{exp}\left(-\dfrac{(E_r-p_3)^2}{2p_4^2}\right) \right]
\end{equation}
where $p_{j=0,4}$ are free parameters: $p_0= 1.34~\mathrm{dru}$, $p_1=  -0.058~\mathrm{keV}^{-1}$, $p_2=  0.2~\mathrm{dru}$, $p_3=  40~\mathrm{keV}$, $p_4=  11.4~\mathrm{keV}$.\\
\noindent  $\bullet$  Recoil energy spectrum for surface $^{206}$Pb recoils is given by  a gaussian distribution associated with a flat component and $p_{j=0,3}$ as free parameters:
\begin{equation}
\label{Eq-surface-206Pb}
\dfrac{dR}{dE_r}=\left[p_0\, +p_1\,\mathrm{exp}\left(-\dfrac{(E_r-p_2)^2}{2p_3^2}\right)\right] 
\end{equation}
$p_0= 0.037~\mathrm{dru}$, $p_1= 0.15~\mathrm{dru}$, $p_2=  95~\mathrm{keV}$, $p_3=  5.7~\mathrm{keV}$.\\
\noindent  $\bullet$  Recoil energy spectrum of heat-only events is parametrized by the sum of two exponential functions:
\begin{equation}
\label{Eq-HO}
\dfrac{dR}{dE_r}=\left[p_0\, \mathrm{exp} \left(-p_1\, E_r\right)+p_2\,\mathrm{exp}\left(-p_3\, E_r \right)\right] 
\end{equation}
where $p_{j=0,3}$ are free parameters: $p_0= 38.2725~\mathrm{dru}$, $p_1=  0.293~\mathrm{keV}^{-1}$, $p_2= 1.4775~\mathrm{dru}$, $p_3=  0.0812~\mathrm{keV}^{-1}$.\\
\noindent  $\bullet$   Spectral shape of radiogenic neutrons consists of the sum of two exponentials:
\begin{equation}
\label{Eq-radiogenic}
\dfrac{dR}{dE_r}=\left[p_0\, \mathrm{exp} \left(-p_1\, E_r\right)+p_2\,\mathrm{exp}\left(-p_3\, E_r \right)\right] 
\end{equation}
with $p_{j=0,3}$ as free parameters: $p_0= 4.827 \times 10^{-4}~\mathrm{dru}$, $p_1=  0.3906~\mathrm{keV}^{-1}$, $p_2= 2.986\times 10^{-4}~\mathrm{dru}$, $p_3=  0.05549~\mathrm{keV}^{-1}$.

\begin{appendix}
\end{appendix}
\end{document}